\documentclass[aps,reprint,showpacs,superscriptaddress]{revtex4-2}

\usepackage{amssymb}
\usepackage{natbib}
\usepackage{graphicx}
\usepackage{amsmath}
\usepackage[bookmarks = false]{hyperref}
\usepackage{bm}
\usepackage[all]{hypcap}
\usepackage{colortbl}
\usepackage{booktabs}
\usepackage{braket}
\usepackage{mathrsfs}
\usepackage{multirow}
\usepackage{upgreek}
\usepackage{textcomp}
\usepackage{soul}
\usepackage{dsfont}
\usepackage[T1]{fontenc}

\usepackage{lineno}

\usepackage{array}
\newcommand{\PreserveBackslash}[1]{\let\temp=\\#1\let\\=\temp}
\newcolumntype{C}[1]{>{\PreserveBackslash\centering}p{#1}}
\newcolumntype{R}[1]{>{\PreserveBackslash\raggedleft}p{#1}}
\newcolumntype{L}[1]{>{\PreserveBackslash\raggedright}p{#1}}

\newcommand{\ustc}{
	\affiliation{Hefei National Laboratory for Physical Sciences at Microscale and Department
	of Modern Physics, University of Science and Technology of China, Hefei,
	Anhui 230026, China}
	\affiliation{CAS Center for Excellence in Quantum Information and Quantum Physics, University of Science and Technology of China, Hefei, Anhui 230026, China}
	\affiliation{Hefei National Laboratory, University of Science and Technology of China, Hefei 230088, China}
}
\newcommand{\delft}{
	\affiliation{Kavli Institute of Nanoscience, Department of Quantum Nanoscience, Delft University of Technology, 2628CJ Delft, The Netherlands}
}
\newcommand{\jinan}{
	\affiliation{Jinan Institute of Quantum Technology, Jinan, China}
}
\newcommand{\anhui}{
	\affiliation{Anhui Provincial Key Laboratory of Photonics Devices and Materials, Anhui Institute of Optical and Fine Mechanics, Hefei Institutes of Physical Science, Chinese Academy of Science, Hefei, Anhui 230031, China}
	\affiliation{Advanced Laser Technology Laboratory of Anhui Province, Hefei, Anhui 230037, China}	
}
\newcommand{\optoelectronic}{
	\affiliation{School of Environmental Science and Optoelectronic Technology, University of Science and Technology of China, Hefei, Anhui 230026, China}
}
\newcommand{\coauthors}{
	\thanks{These authors contributed equally to this work.}
}
\begin{document}

\title{A multinode quantum network over a metropolitan area}
\author{Jian-Long Liu}\coauthors\ustc
\author{Xi-Yu Luo}\coauthors\ustc
\author{Yong Yu}\coauthors\ustc\delft
\author{Chao-Yang Wang}\ustc
\author{Bin Wang}\ustc
\author{Yi Hu}\ustc
\author{Jun Li}\ustc
\author{Ming-Yang Zheng}\jinan
\author{Bo Yao}\anhui
\author{Zi Yan}\ustc
\author{Da Teng}\ustc
\author{Jin-Wei Jiang}\ustc
\author{Xiao-Bing Liu}\anhui
\author{Xiu-Ping Xie}\jinan
\author{Jun Zhang}\ustc
\author{Qing-He Mao}\anhui\optoelectronic
\author{Xiao Jiang}\ustc
\author{Qiang Zhang}\ustc\jinan
\author{Xiao-Hui Bao}\email{xhbao@ustc.edu.cn}\ustc
\author{Jian-Wei Pan}\email{pan@ustc.edu.cn}\ustc

\begin{abstract}
	\normalsize
	Towards realizing the future quantum internet, a pivotal milestone entails the transition from two-node proof-of-principle experiments conducted in laboratories to comprehensive, multi-node setups on large scales. Here, we report on the debut implementation of a multi-node entanglement-based quantum network over a metropolitan area. We equipped three quantum nodes with atomic quantum memories and their telecom interfaces, and combined them into a scalable phase-stabilized architecture through a server node. We demonstrated heralded entanglement generation between two quantum nodes situated 12.5~km apart, and the storage of entanglement exceeding the round-trip communication time. We also showed the concurrent entanglement generation on three links. Our work provides a metropolitan-scale testbed for the evaluation and exploration of multi-node quantum network protocols and starts a new stage of quantum internet research.
\end{abstract}

\maketitle

\begin{figure*}[tp]
	\centering
	\includegraphics[width=0.65\textwidth]{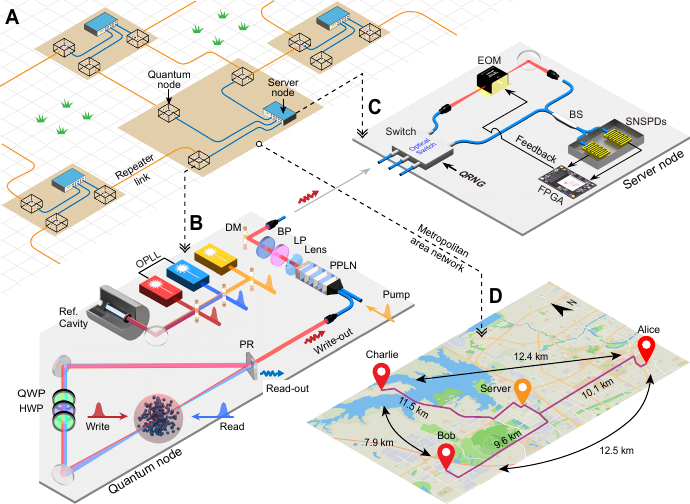}
	 \caption{\textbf{The multinode metropolitan area quantum network.} 
	 (\textbf{A}) An envisioned architecture of a wide-area quantum network. Most of the quantum nodes are densely located in metropolitan areas (shaded in brown), forming numerous metropolitan area networks. Each metropolitan area has a server node that connects and synchronizes all quantum nodes using optical fibers (blue lines), facilitating remote entanglement generation between any pair of quantum nodes upon request. Quantum nodes in different cities establish remote entanglement through quantum repeater links (orange lines) that connect the metropolitan area networks.
	 (\textbf{B}) The scheme of the quantum node. A cloud of $^{87}$Rb atoms in a ring cavity works as a quantum memory. We create local entanglement between the atoms and a write-out photon by applying a write pulse. The write-out photon is collected along the anticlockwise cavity mode and outcoupled through a partially reflective (PR) mirror. The write-out photon is shifted to the telecom wavelength using a PPLN and a 1950~nm pump laser, and then sent to the server node. A combination of dichroic mirrors (DM), long-pass filters (LP), and bandpass filters (BP) is used for noise filtering in the conversion process. The atomic state can be retrieved to the clockwise cavity mode by applying a read laser pulse. All lasers are locked to a reference cavity. The write and read lasers are furthermore synchronized using an OPLL.
	 (\textbf{C}) The scheme of the server node. An optical switch controlled by a QRNG routes two out of three inputs to the interferometer. A beamsplitter (BS) performs single-photon interference, and the results are measured by two SNSPDs. An FPGA is used to analyze the PP interference data from the SNSPDs and compensate the phase to the EOM on one arm of the interferometer (see the main text).
	 (\textbf{D}) The layout of the multinode quantum network studied in this work. Three quantum nodes, labeled Alice, Bob, and Charlie, and a server node are located in four laboratories in the Hefei city. The violet lines represent the fiber connections (lengths indicated), and the black lines indicate the physical distance between quantum nodes. Map data from Mapbox and OpenStreetMap.}

	\label{fig:setup}
\end{figure*}

Quantum networks~\cite{kimble2008,wehner2018} operate differently from their classical counterparts, as their nodes can store and process quantum information and are linked through the sharing of entangled states. The non-local nature of entanglement enables a range of disruptive applications such as quantum cryptography~\cite{gisin2002}, distributed quantum computing~\cite{jiang2007a}, and enhanced sensing~\cite{gottesman2012,komar2014}. Quantum nodes within a metropolitan area can be entangled through optical channels of a few tens of kilometers with a moderate transmission loss (e.g. $10$~dB for 50~km fiber at 1550~nm). For entangling intercity nodes, quantum repeater protocols~\cite{briegel1998,sangouard2011} need to be implemented to overcome the exponential growth in transmission loss. Fig.~\ref{fig:setup}A envisions a wide area complex quantum network, consisting of a number of high-connectivity metropolitan area networks that are furthermore interconnected via quantum repeater channels.

The building block of a quantum network is the heralded entanglement between a pair of adjacent quantum nodes. This has been successfully demonstrated across various physical platforms including atomic ensembles~\cite{chou2005,chou2007,yuan2008}, single atoms~\cite{hofmann2012}, diamond nitrogen-vacancy centers~\cite{bernien2013,hensen2015,humphreys2018}, trapped ions~\cite{moehring2007}, rare-earth doped crystals~\cite{lago-rivera2021,liu2021m} and quantum dots~\cite{delteil2016,stockill2017}, among others. Currently, the maximum physical separation achieved is 1.3~km~\cite{hensen2015}. Building on this progress, three-node quantum network primitives have been established recently on a laboratory scale~\cite{jing2019,pompili2021,hermans2022a}, opening up opportunities for multi-user applications such as the creation of three-body entangled GHZ states~\cite{jing2019,pompili2021}, entanglement swapping between two elementary links~\cite{pompili2021}, and qubit teleportation between non-neighboring nodes~\cite{hermans2022a}. The next challenge is to extend these laboratory or short-scale experiments to a metropolitan scale, which needs the combination of several requirements: low fiber transmission loss, independent quantum nodes and a scalable network architecture. Thanks to the great progress of the quantum frequency conversion (QFC) technique~\cite{kumar1990}, platforms operating at high transmission-loss wavelengths are now able to access the fiber network with minimal loss. This has enabled a series of works distributing local light-matter entanglement over up to a few kilometer fibers~\cite{degreve2012,bock2018,vanleent2020,tchebotareva2019,krutyanskiy2019,luo2022}. Furthermore, two seminal experiments demonstrated heralded entanglement between two quantum nodes via long fiber links, one using two single atoms and 33~km of fiber~\cite{vanleent2022} and the other using two atomic ensembles and 50~km of fiber~\cite{yu2020}. However, the entanglement between two single atoms was generated at a low rate of approximately 1~min$^{-1}$ and the entanglement between two atomic ensembles was generated between two non-independent nodes, raising concerns about their scalability towards large-scale and multi-user scenarios.

Here, we overcome these challenges with a set of key innovations and present a multinode quantum network over a metropolitan area. Our network consists of three quantum nodes (Alice, Bob, and Charlie) and a server node arranged in a star topology (Fig.~\ref{fig:setup}D). The three quantum nodes are located on the three vertices of a triangle with sides of 7.9~km to 12.5~km in length in Hefei city, and run independently. The server node is situated approximately in the center of the triangle and linked to each quantum node via optical fibers for both classical and quantum communication. In each quantum node, a cavity enhanced laser-cooled Rubidium atomic ensemble serves as the long-lived quantum memory~\cite{bao2012}, generating atom-photon entanglement using the Duan-Lukin-Cirac-Zoller (DLCZ) protocol~\cite{duan2001}. The photon is sent to the server node for entangling, while the atomic qubit is stored for subsequent applications. To reduce photon loss in the fiber, each quantum node is equipped with a QFC module, which coherently shifts visible photons at the Rubidium resonance to the telecom O band~\cite{yu2020}. Quantum nodes are furthermore synchronized in phase using a remote phase stabilization method. Using this architecture, we demonstrated the generation of entanglement between distant quantum nodes and the storage of entanglement exceeding the round-trip communication time. Furthermore, we extended the remote entanglement generation to all three links in the network and executed them concurrently.

\section*{Network architecture for remote entanglement generation}

\begin{figure*}[htbp]
	\centering
	\includegraphics[width=0.65\textwidth]{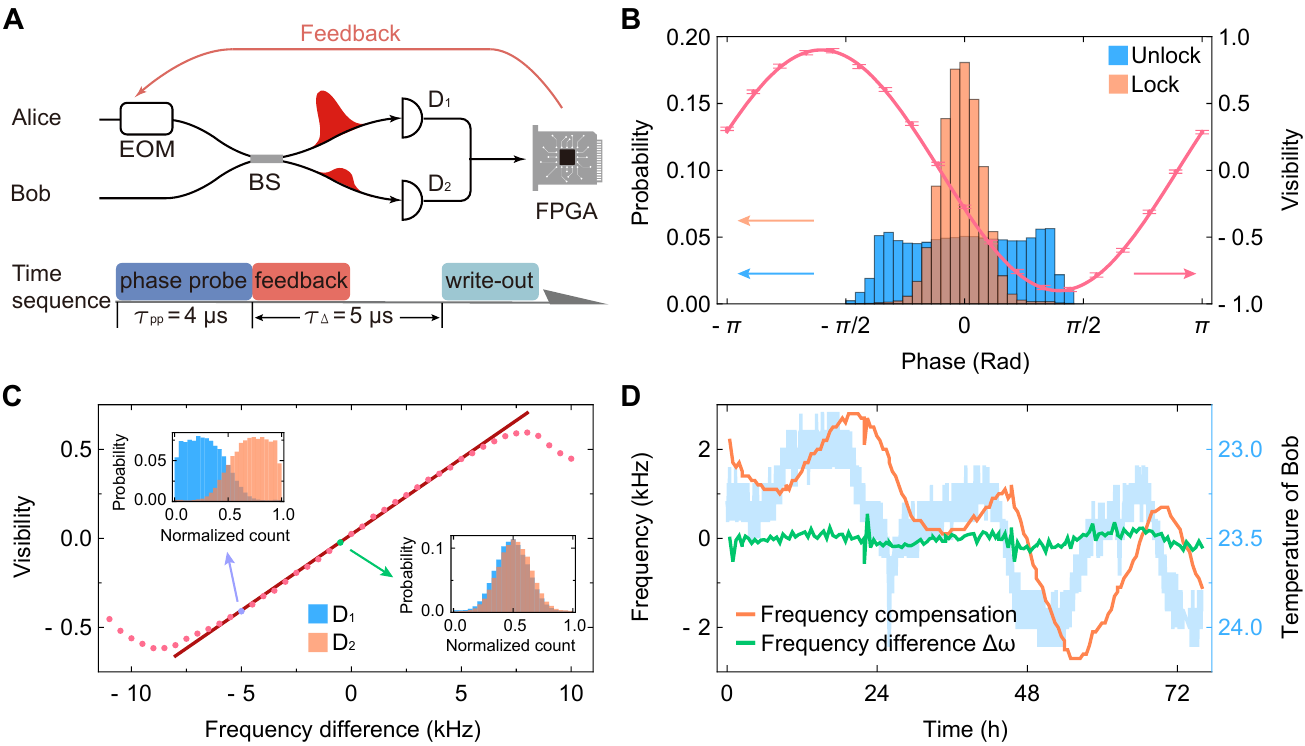}
	\caption{\textbf{Weak-field phase and frequency stabilization.} 
	(\textbf{A}) Schematic and time sequence of the phase stabilization method. We probe the phase difference of two nodes by interfering two PP pulses on the entanglement generation facilities. A weak pulse intensity is used to prevent SNSPD latching. An FPGA extracts the phase difference by analyzing the counts of two SNSPDs and applies the feedback to the EOM. In each trial, PP pulses are sent before the write process, and a gap of $\tau_{\Delta}=5~\upmu$s is secured to reduce the afterpulses-like noise of SNSPDs. Each PP pulse lasts for $\tau_{\textrm{pp}}=4~\upmu$s and contains $\sim100$ photons upon arrival at the beamsplitter. The feedback takes up to 2~$\upmu$s. In practice, we execute an additional probe-feedback right before the stabilization process to improve performance.
	(\textbf{B}) Characterization of the phase stabilization. Two histograms show the detected phase distributions with (orange) and without (blue) executing the phase stabilization. When running the phase stabilization, we perform single-photon interference by sending weak pulses from two nodes. The pink data points show the interference visibility as a function of their phase difference. The pink curve is a sinusoidal fitting of the data points. The error bars represent one standard deviation.
	(\textbf{C}) Interference visibility of PP pulses as a function of the frequency difference between two nodes. Red data points show the variation in visibility when deliberately changing the frequency difference. The red line is a linear fitting of the data points within $\pm 5$~kHz. The two histograms in insets show the distribution of the counts of two SNSPDs (D$_1$ and D$_2$) within a 120~s timeslot, with the corresponding data points highlighted.
	(\textbf{D}) Tracking of laser drift over 76 hours. The green curve shows the detected frequency difference after compensation. The orange curve plots the compensation value, which approximates the original laser drift. We plot the temperature of Bob in blue, which shows a strong correlation with the laser drift.}
	\label{fig:phase detail}
\end{figure*}

We generate remote entanglement between a pair of distant quantum nodes via a single photon scheme~\cite{duan2001}. We start by preparing the light-matter entanglement in each quantum node (Fig.~\ref{fig:setup}B) using the DLCZ protocol~\cite{duan2001}. A weak write pulse induces a spontaneous Raman-scattered photon (write-out) in a small probability $\chi\approx1\%$, whose presence and absence is associated to that of collective excitation of the atomic ensemble. This forms a Fock-basis light-matter entanglement $\ket{\Phi_{\textrm{ap}}}=\ket{00}+\sqrt{\chi}\ket{11}$, where $0$ and $1$ refer to the number of photon or atomic excitation, indicated by the subscript a and p, respectively, from hereafter. The atomic excitation can be stored up to 560~$\upmu$s until being retrieved for following operations~\cite{luo2022}. The write-out photon is down-converted from 795~nm (3.5~dB/km) at Rubidium resonance to 1342~nm (0.3~dB/km) in the telecom O band in a periodically poled lithium niobate (PPLN) waveguide chip with the help of a strong 1950~nm Pump laser. The telecom photon then is sent to the server node for interference. The QFC module, including the down-conversion, noise filtering and fiber coupling, constitute an end-to-end efficiency of 46\% and noise of $\sim100$~Hz. In the server node (Fig.~\ref{fig:setup}C), a beamsplitter combines the write-out photon from two quantum nodes and erases their which-path information. A click from a superconducting nanowire single photon detector (SNSPD) heralds the maximally entangled states between two ensembles 
\begin{equation}\label{eq:EntFock}
	\ket{\Psi_{\textrm{aa}}^{\pm}}=(\ket{01}\pm e^{i\Delta\varphi}\ket{10})/\sqrt{2},
\end{equation}
where the $\pm$ sign depends on the heralding detector. $\Delta\varphi=\Delta\varphi_\text{w} +\Delta\varphi_\text{p} + \Delta\varphi_{\text{f}}$ accounts for the optical phase difference accumulated during the entangling process, including $\Delta\varphi_\text{w}$ and $\Delta\varphi_\text{p}$, the phase difference of Write lasers and of Pump lasers in two nodes, respectively, as well as $\Delta\varphi_{\text{f}}$, the phase difference of two fiber channels.

The key to the success of the single-photon scheme lies in accurately setting $\Delta\varphi$ to a known value. To this end, previous laboratory experiments have typically used an ad hoc laser shared by separated nodes, and locked the closed-loop interferometer composed of the laser splitting paths and the quantum channels~\cite{chou2005,stockill2017,yu2020,pompili2021,lago-rivera2021}. It adds complexity to the networking by requiring additional fiber and hardware resources and exposes the quantum nodes to the risk of potential attacks during quantum communication~\cite{pang2020a}. Crucially, additional unknown fiber phases are introduced to the generated state, which erases the state's coherence.

To address these challenges, we developed a network architecture based on a weak-field remote phase stabilization method. The method relies on detecting the phase difference $\Delta\varphi$ in situ at the entanglement generation facilities using weak-field phase probe (PP) pulses and photon counting, and compensating for it immediately, similar to that being realized recently in quantum key distribution experiments~\cite{zhou2023}. To obtain an accurate measurement of $\Delta\varphi$, we prepare a PP pulse with a length of $\tau_{\textrm{pp}}=4~\upmu$s at the write-out photon's frequency from the Read laser (due to its frequency likeness to the write-out photon) in each node, and use it to simulate the write-out photon's transmission $\tau_{\Delta}=5~\upmu$s earlier than the entanglement trial. Analogous to the entangling process, two PP pulses interfere in the server node, with the results reflected in the counts of SNSPDs. We extract the phase difference $\Delta\varphi_{\text{pp}}$ from the interference and compensate for it in an electro-optical modulator (EOM) before the interference beamsplitter. This method has three critical requirements for the PP pulses: their frequency difference is much smaller than $1/\tau_{\textrm{pp}}$, their linewidths are much narrower than $1/\tau_{\textrm{pp}}$ and $1/\tau_{\Delta}$, and they inherit the same laser phase as the write-out photons. To meet these requirements, we lock all lasers in each node onto a reference cavity and get a linewidth of sub-kilohertz and a drift of a few kilohertz per day for all lasers. We perform a one-off frequency calibration of the corresponding lasers in different nodes. Additionally, we synchronize the Write and Read lasers using an optical phase-locked loop (OPLL). As a result, each PP pulse differs from the corresponding write-out photon with a phase $\theta$, leading to a residual phase $\Delta\theta$ (the difference of $\theta$ in two nodes) in Eq.~\ref{eq:EntFock} after compensation. One can either remove $\Delta\theta$ by synchronizing the OPLLs in two nodes or choose $\Delta\theta$ as the phase reference to observe the entanglement. We chose the latter method in the following experiments.

In practice, we conduct twice the phase stabilization in each trial for a better performance (Fig.~\ref{fig:phase detail}A). To characterize the stabilization performance, we send an additional PP pulse from each quantum node 5~$\upmu$s after the feedback and find a Gaussian phase distribution with a standard deviation of 17$^{\circ}$, as shown in Fig.~\ref{fig:phase detail}B. By changing the relative phase between the additional pair of PP pulses, we observe a sinusoidal oscillation of their interference visibility (Fig.~\ref{fig:phase detail}B), which confirms a good phase correlation between two nodes.

An additional challenge in operations arises from the slow frequency drift of the reference cavities, which can transfer to the lasers and eventually to write-out photons. To avoid this, we employ a continuous frequency calibration using the PP interference data. We determine the minor laser frequency offset $\Delta \omega$ by analyzing the phase evolution between successive entanglement trials over a duration of $t$ and compensate it to the Pump lasers. Considering the long coherence time of the lasers ($\approx 1$~ms) and the slow fluctuations in the fiber, the phase evolves solely due to $\Delta\omega t$. We verify this by observing the interference visibility as a function of $\Delta \omega$ shown in Fig.~\ref{fig:phase detail}C. Through the implementation of the calibration, we are able to maintain $\Delta\omega\leq 0.1$~kHz over a span of 76 hours, despite an observed drift of 5~kHz (Fig.~\ref{fig:phase detail}D).

\section*{Entanglement between a pair of distant nodes}
\begin{figure*}
	\centering
	\includegraphics[width=0.65\textwidth]{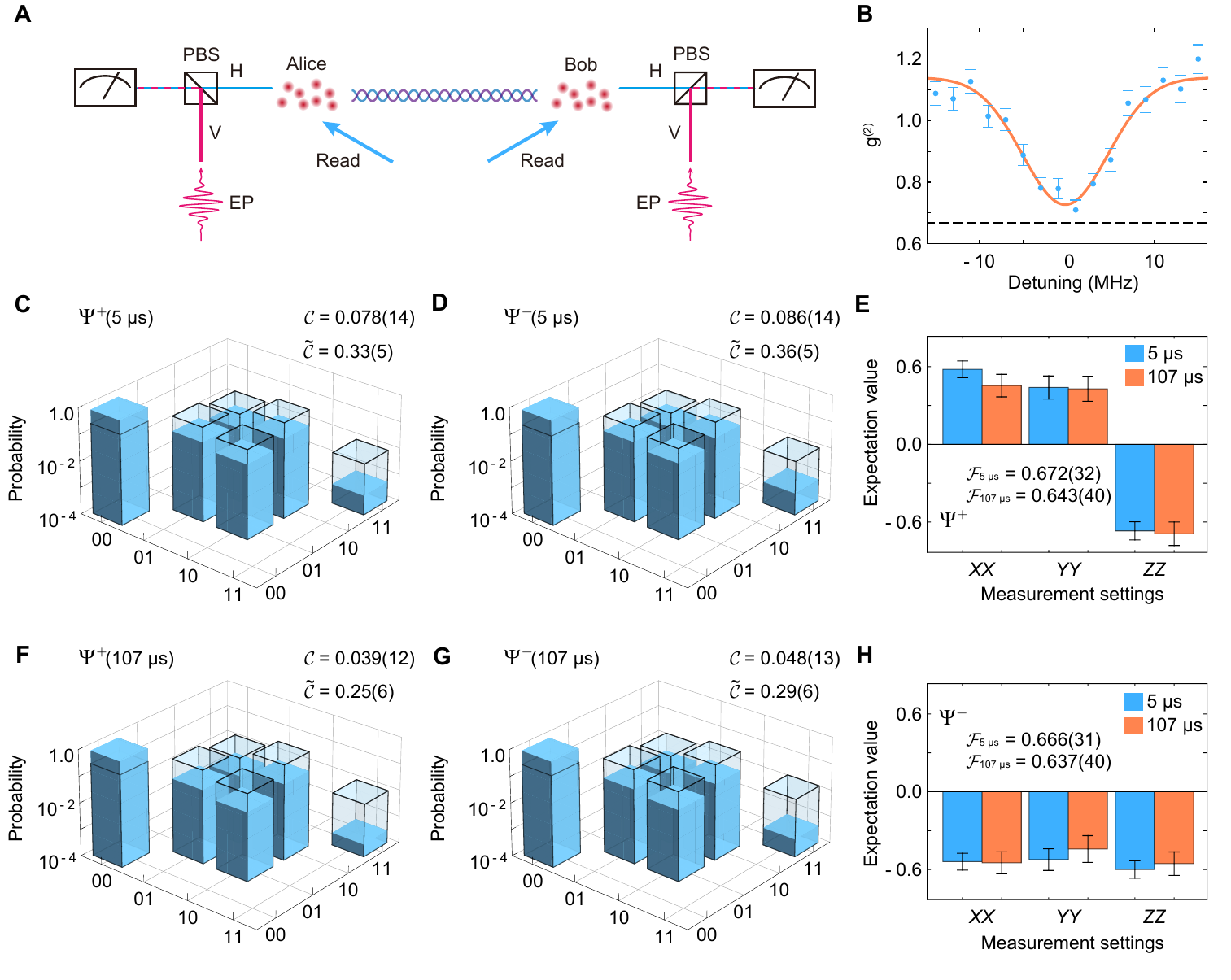}
	 \caption{\textbf{Entanglement between a pair of distant quantum nodes.} 
	 (\textbf{A}) The experimental setup for entanglement verification. In each node, a weak EP pulse that has the same frequency and profile as the read-out field but is orthogonally polarized assists in the verification process. We combine the EP pulses and the read-out fields using PBSs, perform the polarization measurement in each node, and detect the remote entanglement by analyzing the correlation between two nodes (see the main text).
	 (\textbf{B}) Measured correlation function $g^{(2)}$ (results of the Hong-Ou-Mandel experiment) between a read-out field and an EP pulse as a function of their detuning. The orange line shows a Gaussian fit and the black dashed line indicates the thermal-coherent interference limit.
	 (\textbf{C}, \textbf{D}) Reconstructed density matrices of $\ket{\Psi_{\textrm{pp}}^+}$ and $\ket{\Psi_{\textrm{pp}}^-}$, respectively, after 5~$\upmu$s of storage. 
	 (\textbf{F}, \textbf{G}) Reconstructed density matrices of $\ket{\Psi_{\textrm{pp}}^+}$ and $\ket{\Psi_{\textrm{pp}}^-}$, respectively, after 107~$\upmu$s of storage. In the density matrix figures, blue bars refer to the tomography of the read-out fields, and the transparent bars rescale the tomography to two atomic ensembles.  
	 (\textbf{E}, \textbf{H}) Correlation measurements of $\ket{\Psi_{\textrm{PME}}^+}$ and $\ket{\Psi_{\textrm{PME}}^-}$, respectively. The blue and orange bars refer to the results for the 5~$\upmu$s and 107~$\upmu$s storage cases, respectively. The error bars represent one standard deviation.}
	\label{fig:two node result}
\end{figure*}

We first generate remote entanglement between Alice and Bob. In addition to laser frequency and phase stabilization, we apply polarization filtering using a fiber polarizing beamsplitter (PBS) for the photons when they arrive the server node and continuously adjust an electrical polarization controller before the PBS to maximize the filtering efficiency (not shown in the setup schematic). To measure the distant entanglement between the two atomic ensembles, we retrieve each atomic state to a photonic mode via a read pulse generated from the Read laser, and obtain $\ket{\Psi_{\text{pp}}^{\pm}}=(\ket{01}\pm e^{i\Delta\varphi_r}\ket{10})/\sqrt{2}$. To evaluate the entanglement, one needs the knowledge of coherence $d$ between $\ket{01}$ and $\ket{10}$. As it is experimentally hard to perform qubit operations in the Fock basis, prior experiments mostly inferred $d$ by interfering two read-out modes using a beamsplitter~\cite{chou2005,yu2020,lago-rivera2021}, which is not favorable for spatially separate cases. To overcome this challenge, we develop a method based on the weak-field homodyne measurement schemes~\cite{tan1991,li2013} that allows us to witness remote Fock-basis entanglement through local operations and measurements. Additionally, our method effectively converts the Fock-basis entanglement into a polarization maximally entangled (PME) state~\cite{duan2001}, making it accessible for entanglement-based communication schemes.

Our method starts by generating an entanglement probe (EP) pulse from the Write laser in each node. It is a weak coherent light field that has the same frequency and profile as the read-out field, but orthogonally polarized. When choosing the phase reference of $e^{i\Delta\theta}$, one can write the joint state of two EP pulses as $\ket{\textrm{EP}}_{\textrm{A}}\ket{\textrm{EP}}_{\textrm{B}}=|\alpha|_A|\alpha|_B e^{i\Delta\varphi_r}$, which is in phase with $\ket{\Psi_{\text{pp}}^{\pm}}$. Next, we combine the read-out field with the EP pulse in every node using a PBS and perform projection measurements under different polarization bases (Fig.~\ref{fig:two node result}A). Initially, we measure the mixed field in both nodes in the $\ket{H}/\ket{V}$ basis. On the detector corresponding to horizontal polarization, the EP is removed, thus we find $p_{00}$, $p_{01}$, $p_{10}$, and $p_{11}$ through statistical analysis, where $p_{ij}$ represents the probability of having $i$ excitations in Alice and $j$ excitations in Bob. When measuring in the superposition basis, the read-out field interfere with the EP pulses, resulting in remote correlation. Specifically, when $\ket{H}\pm\ket{V}$ basis is chosen, we have the correlation function
\begin{equation}
E_{\pm}=\frac{N_{++}+N_{--}-N_{+-}-N_{-+}}{N_{++}+N_{--}+N_{+-}+N_{-+}},
\end{equation}
where $N_{mn}$ represent the coincidence events between the $m$ mode in Alice and $n$ mode in Bob ($m,n$ takes $+$ or $-$, referring to $\ket{H}+\ket{V}$ and $\ket{H}-\ket{V}$, respectively). In this case, we find $|d|\geq (p_{01}+p_{10}+2\sqrt{p_{00}p_{11}})|E_{\pm}|/2$. Now we can estimate the degree of entanglement with concurrence $\mathcal{C}=\textrm{max}(0, 2|d|-2 \sqrt{p_{00}p_{11}})$. This measure ranges from 0 for a separable state to 1 for a maximally entangled state. 

Although being a good measure of entanglement, $\mathcal{C}$ does not indicate how useful the entangled state is in entanglement-based communication. To answer this question, we consider the mixed field right after the combining PBS at each node, which can be approximated (with arbitrary phase factors and no normalization)
\begin{equation}\label{eq:CombineState}
	(a^{\dagger}_{\textrm{A},H} \pm a^{\dagger}_{\textrm{B},H}) \otimes (\mathds{1}+\alpha a^{\dagger}_{\textrm{A},V}) \otimes (\mathds{1}+\alpha a^{\dagger}_{\textrm{B},V}) \ket{0},
\end{equation}
where $\mathds{1}$ is the identity operator and $a^{\dagger}$ is the photon creation operator with the subscript indicating its node (Alice or Bob) and polarization ($H$ or $V$). Selecting at least one photon at each node, Eq.~\ref{eq:CombineState} becomes
\begin{equation}\label{eq:PME}
	\ket{\Psi_{\textrm{PME}}^{\pm}}=\alpha\left(a^{\dagger}_{\textrm{A},H} a^{\dagger}_{\textrm{B},V} \pm a^{\dagger}_{\textrm{A},V} a^{\dagger}_{\textrm{B},H}\right)\ket{0} + o(\alpha^2).
\end{equation}
Neglecting the higher-order term $o(\alpha^2)$, Eq.~\ref{eq:PME} is a PME state between two photonic modes. This allows local Pauli operations ($X$, $Y$, and $Z$) and enables entanglement based communication protocols. More details about the weak-field entanglement verification method are discussed in Ref.~\cite{SM}.

The weak-field method to verify remote entanglement relies on the EP pulse's likeness to the read-out field. We examine this by performing a Hong-Ou-Mandel (HOM) experiment between these two fields. By scanning the detuning of the EP pulse, we observe a HOM dip in the correlation function $g^{(2)}$ shown in Fig.~\ref{fig:two node result}B, and get a minimum value of $0.716(19)$, close to the thermal-coherent interference limit of $2/3$ (the EP pulse is a coherent state and the read-out field without conditioning by the write-out field is a thermal state). This will lead to a 10\% underestimation of $d$ in measurements~\cite{SM}.

Next, we verify the remote entanglement generated between Alice and Bob using the weak-field method. In the first experiment, we verify the entanglement in a delayed-choice fashion~\cite{ma2012a}. The atomic states are retrieved 5~$\upmu$s after the atom-photon entanglement is created, earlier than the entanglement is heralded. Fig.~\ref{fig:two node result}C and D summarize the reconstructed density matrices between the read-out fields and the corresponding $\mathcal{C}$. $\mathcal{C}>0$ for both $\ket{\Psi^+_{\textrm{pp}}}$ and $\ket{\Psi^-_{\textrm{pp}}}$ witness the existence of remote entanglement. By deducting the loss during retrieval and detection, we infer the density matrices between two atomic ensembles and their corresponding concurrence $\tilde{\mathcal{C}}$ to two atomic ensembles and show them in the same figures. Meanwhile, we find the fidelity $\mathcal{F}=0.672(32)$ for $\ket{\Psi_{\textrm{PME}}^{+}}$ and 0.666(31) for $\ket{\Psi_{\textrm{PME}}^{-}}$ to corresponding maximally entangled state, both above 0.5 to certify the entanglement (Fig.~\ref{fig:two node result}E and H). In the second experiment, we create the remote entanglement in the same fashion but store it for $107~\upmu$s, exceeding the round-trip transmission time between the quantum and the server nodes. We get a slightly lower concurrence result for each case, as summarized in Fig.~\ref{fig:two node result}F and G. However, the fidelities $\mathcal{F}$ of $\ket{\Psi_{\textrm{PME}}^{\pm}}$ are on par with that of the former experiment (Fig.~\ref{fig:two node result}E and H). Comparing the variation of $\mathcal{C}$ and $\mathcal{F}$ between two experiments, we confirm that a large portion of noise in the generated entangled states is the vacuum component and can be effectively filtered out using our weak-field method and other protocols~\cite{duan2001}. The measured heralding probability is $6.9\times 10^{-4}$ and $8.0\times 10^{-4}$ for the first and second experiments, respectively, which corresponds to an entangling rate of 1.93~Hz and 0.83~Hz, considering the experiment repetition rate of 2.76~kHz and 1.03~kHz.

\section*{Concurrent entanglement generation in the network}

\begin{figure*}[htbp]
	\centering
	\includegraphics[width=0.65\textwidth]{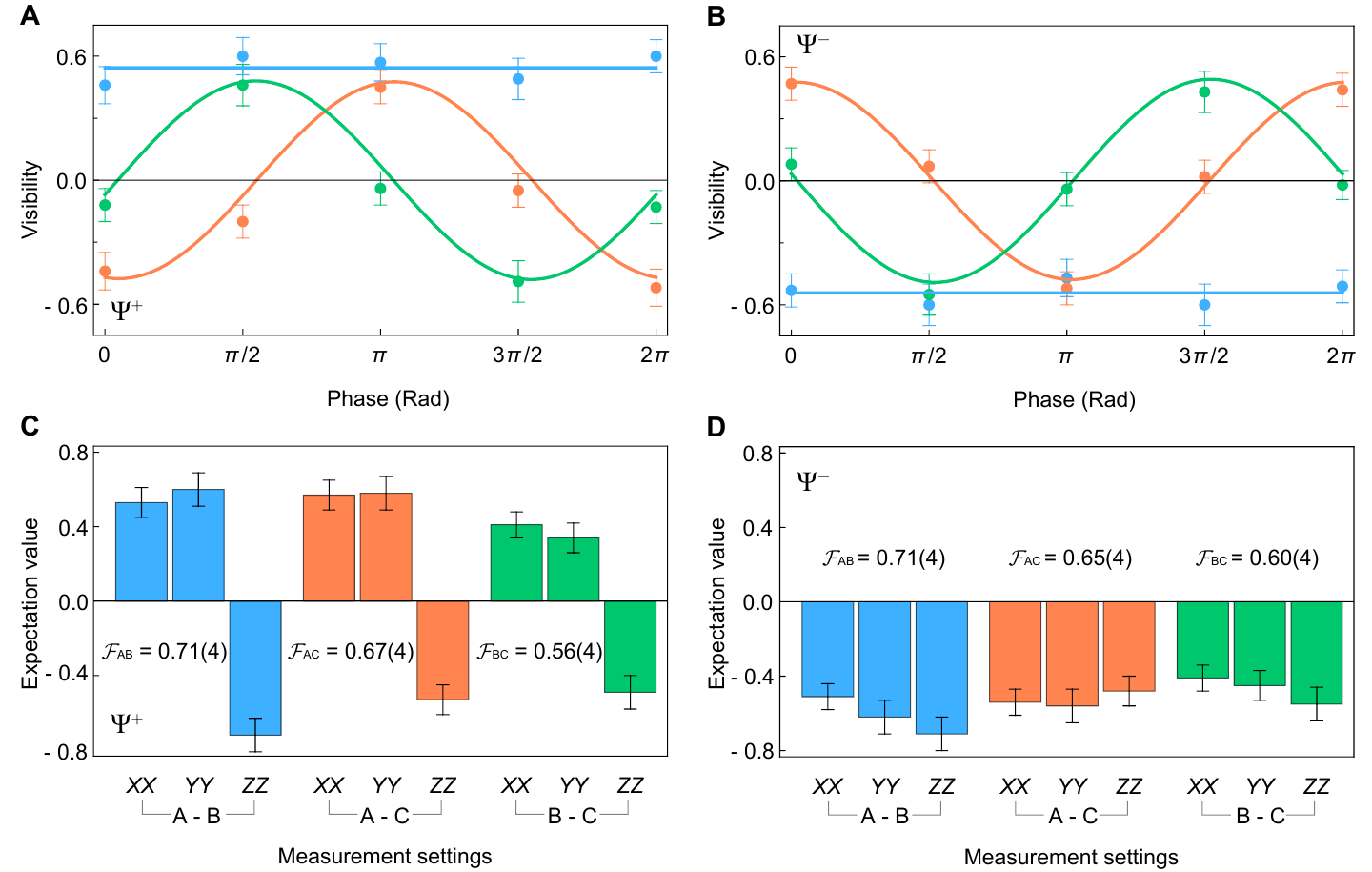}
	 \caption{\textbf{Concurrent entanglement generation in the network.} 
	 (\textbf{A}, \textbf{B}) Measured correlation $\braket{XX}$ of $\ket{\Psi^+}$ and $\ket{\Psi^-}$, respectively, on three links when changing the phase $\theta$ at Charlie. Blue, orange, and green data points refer to the results of the Alice-Bob, Alice-Charlie, and Bob-Charlie links, respectively. Curves show the sinusoidal or the linear fit of data with corresponding colors.
	 (\textbf{C}, \textbf{D}) Correlation measurement of $\ket{\Psi_{\textrm{PME}}^+}$ and $\ket{\Psi_{\textrm{PME}}^-}$, respectively, on three links. The color code remains the same as in (A) and (B). The error bars represent one standard deviation.}
	\label{fig:three node result}
\end{figure*}

\begin{table*}[htbp]
	\centering
	\caption{\textbf{Summary of the concurrence results in the concurrent entanglement generation experiment.}}
	\begin{tabular}[t]{C{0.6cm} C{1cm} C{2.2cm} C{2.2cm} C{2.2cm}}
	\toprule
	& & Alice-Bob  &Alice-Charlie& Bob-Charlie\\
	\midrule
	\multirow {2}*{${\mathcal{C}}$}& ${\ket{\Psi^+}}$&0.086(14)&0.074(14)&0.032(12)\\
	& ${\ket{\Psi^-}}$&0.086(14)&0.063(13)&0.048(13)\\[1.2ex]

	\multirow {2}*{${\tilde{\mathcal{C}}}$} & ${\ket{\Psi^+}}$&0.38(5)&0.35(5)&0.17(5)\\
	& ${\ket{\Psi^-}}$&0.38(5)&0.30(5)&0.23(5)\\
	\bottomrule
	\end{tabular}
	\label{tab:concurrence}
\end{table*}

Now we complete the network architecture by including the third quantum node, Charlie, and concurrently generate entanglement $\ket{\Psi^{\pm}}_{\textrm{AB}}$, $\ket{\Psi^{\pm}}_{\text{BC}}$ and $\ket{\Psi^{\pm}}_{\text{AC}}$ on three links, Alice-Bob, Bob-Charlie and Alice-Charlie, respectively. To this end, the server node is equipped with a multi-input-two-output optical switch and assigns two of the three users to the detection upon request. In our demonstration, we assign a 5.8~s time slot for each pair of users and dynamically switch user pairs according to a quantum random number generator (QRNG), simulating unknown requests in real-world applications. Since different links are frequently switched, a `hot swapping' feature is crucial, i.e. the entangling conditions are meet immediately after switching. In our architecture, the phase stabilization is executed for every entanglement trial and is unaffected by switching. The good laser stability guarantees ignorable relative frequency drift although a 1/3 effective calibration time is assigned for each pair of nodes. Furthermore, we store the optimal EPC parameters for the polarization filtering to each specific fiber and load them during switching.

We first observe the local manipulation of non-local states. For instance, when applying a unitary operation locally at one of the quantum nodes (e.g., Charlie), its relevant non-local states ($\ket{\Psi^{\pm}}_{\text{AC}}$ and $\ket{\Psi^{\pm}}_{\text{BC}}$) change in form while the entanglement is preserved. The irrelevant non-local states $\ket{\Psi^{\pm}}_{\textrm{AB}}$ remain unchanged. In practice, we vary the phase $\theta$ at Charlie and measure the $\braket{XX}$ correlation of each pair of entanglement. As shown in Fig.~\ref{fig:three node result}A and B, we observe sinusoidal oscillations in the expected pairs $\ket{\Psi^{\pm}}_{\text{BC}}$ and $\ket{\Psi^{\pm}}_{\text{AC}}$, while $\ket{\Psi^{\pm}}_{\text{AB}}$ contributes a constant in this measurement. A $\pi/2$ delay between the oscillations of $\ket{\Psi^{\pm}}_{\text{BC}}$ and $\ket{\Psi^{\pm}}_{\text{AC}}$ arises from the phase stabilization mechanism~\cite{SM}.

In the end, we characterize the generated remote entangled states on three links in the delayed-choice fashion. Tab.~\ref{tab:concurrence} summarizes the concurrence results $\mathcal{C}$ and $\mathcal{\tilde{C}}$ and Fig.~\ref{fig:three node result}C, D show the correlation measurement results of the PME states. For all three links, we find $\mathcal{C}>0$ to confirm the presence of atomic entanglement and $\mathcal{F}>0.5$ to witness the entanglement in corresponding PME state. The entangled states between Alice and Bob are comparable to that generated in the fixed connection case, indicating no extra decoherence being introduced in the switching process. The lower entanglement quality in other links arises from experimental imperfections~\cite{SM}. The entangling rates of all three links are roughly 1/3 of that measured in the fixed connection case as expected. Some simple upgrades to the server node will immediately improve the networking performance, such as parallel (rather than concurrent) creation of multiple entanglement pairs with the help of wavelength division multiplexing technique~\cite{wengerowsky2018} and genuine three-body entanglement creation using a modified interferometer~\cite{choi2010,jing2019,dur2000}.

\section*{Discussion and outlook}

In this work, we present the debut implementation of a metropolitan area entanglement-based quantum network. With this facility, we have successfully demonstrated the heralded generation of remote entanglement between two quantum nodes separated by 12.5~km and store it longer than the two-way communication time. We also showed the concurrent generation of multiple entanglement pairs within the network. The high entangling rate in our network is enabled by a series of methods, two of which are crucial. Firstly, the remote phase stabilization guarantees the basis to implement the single photon scheme. In addition, it provides a flexible and scalable network infrastructure that can accommodate more quantum nodes within the metropolitan area. Secondly, the weak-field method simplifies the verification of remote Fock-basis entanglement. It removes the necessity to interfere two remote Fock-basis photons, instead only requires local linear optics operations and measurements. Moreover, using this method, one can effectively convert Fock-basis entanglement to polarization maximally entanglement, making entanglement-based communication within the reach. Adapting these methods to other physical platforms will accelerate their networking process.

Further advances of the network performance will happen in four main aspects, increasing the entangling rate by shifting the wavelength to telecom C band~\cite{vanleent2020,vanleent2022} and by adapting multiplexed atomic quantum memories~\cite{collins2007,pu2017,parniak2017}; decreasing the entanglement infidelity by suppressing unwanted higher order atomic excitations with the help of Rydberg blockade mechanism~\cite{li2013,sun2022}; realizing a better entanglement verification by filtering out the vacuum components in the remote entangled state~\cite{duan2001,chou2007} or by using nonlinear atomic state detection techniques~\cite{xu2021,yang2022d}; and prolonging the entanglement storage time to the second scale by further limiting atomic thermal motion with optical lattice technique~\cite{yang2016,wang2021e}. With these improvements, we expect remote entanglement can be generated much faster than it is lost, laying the basis for mediating serval metropolitan area quantum networks via quantum repeater protocols~\cite{briegel1998,sangouard2011}.

\section*{Acknowledgment}
We acknowledge QuantumCTek for the allocation of node Bob, and Hefei Institutes of Physical Science for the allocation of node Charlie. \textbf{Funding}: This research was supported by the Innovation Program for Quantum Science and Technology (No.~2021ZD0301104), National Key R\&D Program of China (No.~2020YFA0309804), Anhui Initiative in Quantum Information Technologies, National Natural Science Foundation of China, and the Chinese Academy of Sciences. \textbf{Competing interests}: The authors declare no competing interests. \textbf{Data and materials}: The data are archived at Zenodo~\cite{dataset}.

\setcounter{figure}{0}
\setcounter{table}{0}
\setcounter{equation}{0}

\onecolumngrid

\global\long\def\theequation{S\arabic{equation}}
\global\long\def\thefigure{S\arabic{figure}}
\global\long\def\thetable{S\arabic{table}}
\renewcommand{\arraystretch}{0.6}

\newpage

\newcommand{\msection}[1]{\vspace{\baselineskip}{\centering \textbf{#1}\\}\vspace{0.5\baselineskip}}

\msection{SUPPLEMENTAL MATERIAL}

\section{Details of the experimental setup}

\subsection{General details}
We cool down and trap the atomic ensemble using a magneto-optical trap (MOT). We run the MOT at a repetition rate of $\sim 34.5$~Hz. In each MOT cycle, we cool down the atoms in the first 25~ms/24~ms (for a storage time of 5~$\upmu$s/107~$\upmu$s), pump the atoms to the specific Zeeman ground state in the next 1~ms, and perform the entanglement measurement using the rest time. Each entanglement trial takes 33~$\upmu$s in the delayed-choice measurements and 131~$\upmu$s when storing the qubit longer than the round-trip communication time, leading to an average experimental repetition rate of 2.76~kHz and 1.03~kHz.

\begin{figure}[hbtp]
	\centering
	\includegraphics[width=.9\linewidth]{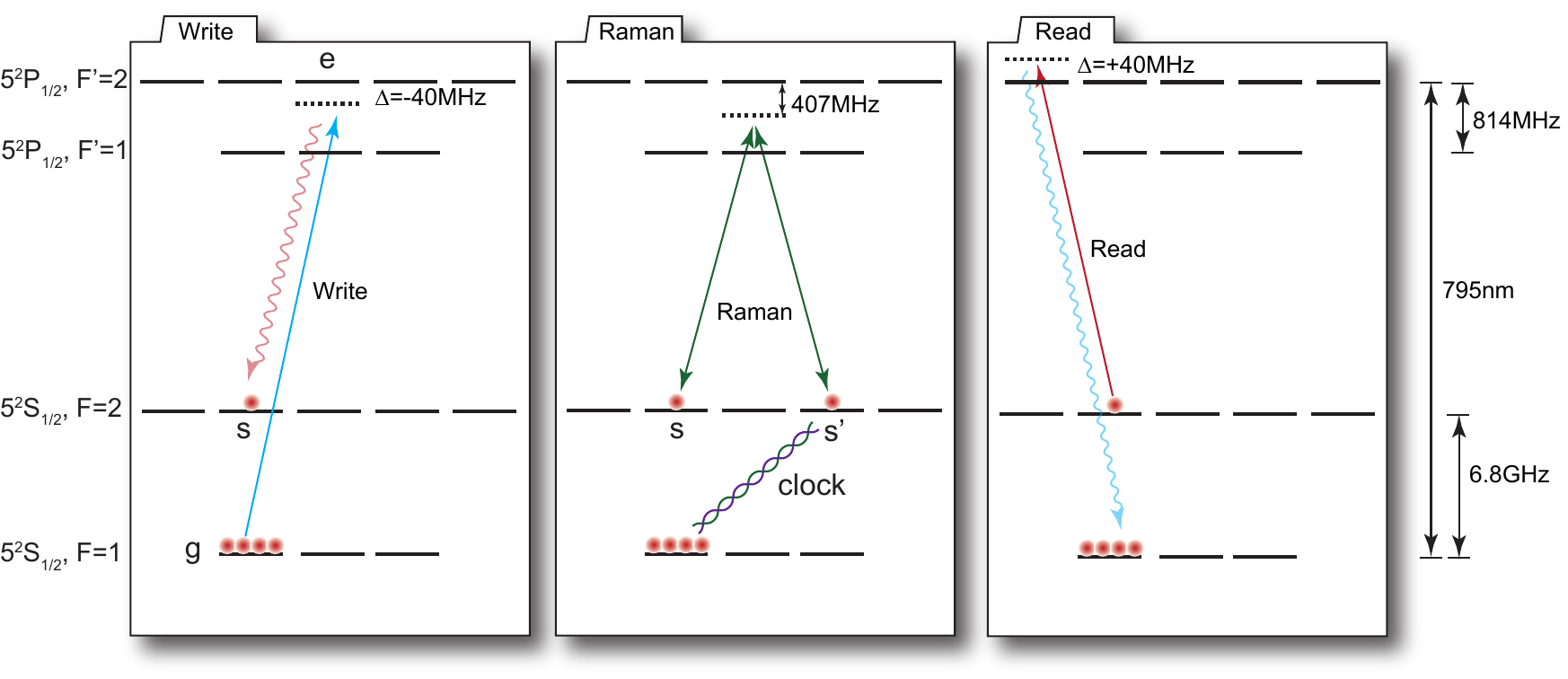}
	\caption{Energy level scheme of main steps at each memory node.}
	\label{fig:energylevel}
\end{figure}

Fig.~\ref{fig:energylevel} depicts the energy level scheme of three main steps at each quantum node, i.e., the write, read process of the DLCZ quantum memory and the spin-wave freezing technique using the Raman operations. More details of the energy level and the spin-wave freezing technique are discussed in Ref.~\cite{luo2022sm}.

\subsection{Experimental control and synchronization}
\begin{figure}[htbp]
	\centering
	\includegraphics[width=.9\linewidth]{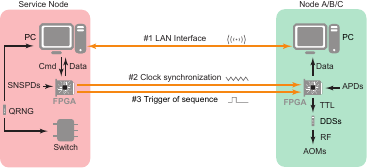}
	\caption{Experimental control and synchronization.}
	\label{fig:controlsystem}
\end{figure}

Fig.~\ref{fig:controlsystem} depicts the schematic of our experimental control system. In each network node, a field programmable gate array (FPGA) takes charge of time sequence generation and data acquisition. The acquiesced data are instantaneously uploaded to a computer for storage. The FPGA in the server node disseminates its clock signal and experimental triggers to the quantum nodes via optical fibers for experiment synchronization. In the server node, we store the random number taken from the quantum random number generator~\cite{nie2015sm} for data synchronization. More details of the system control and synchronization can be found in our previous paper~\cite{luo2022sm}.

\section{The phase of entanglement states}
\begin{figure}[htbp]
	\centering
	\includegraphics[width=.9\linewidth]{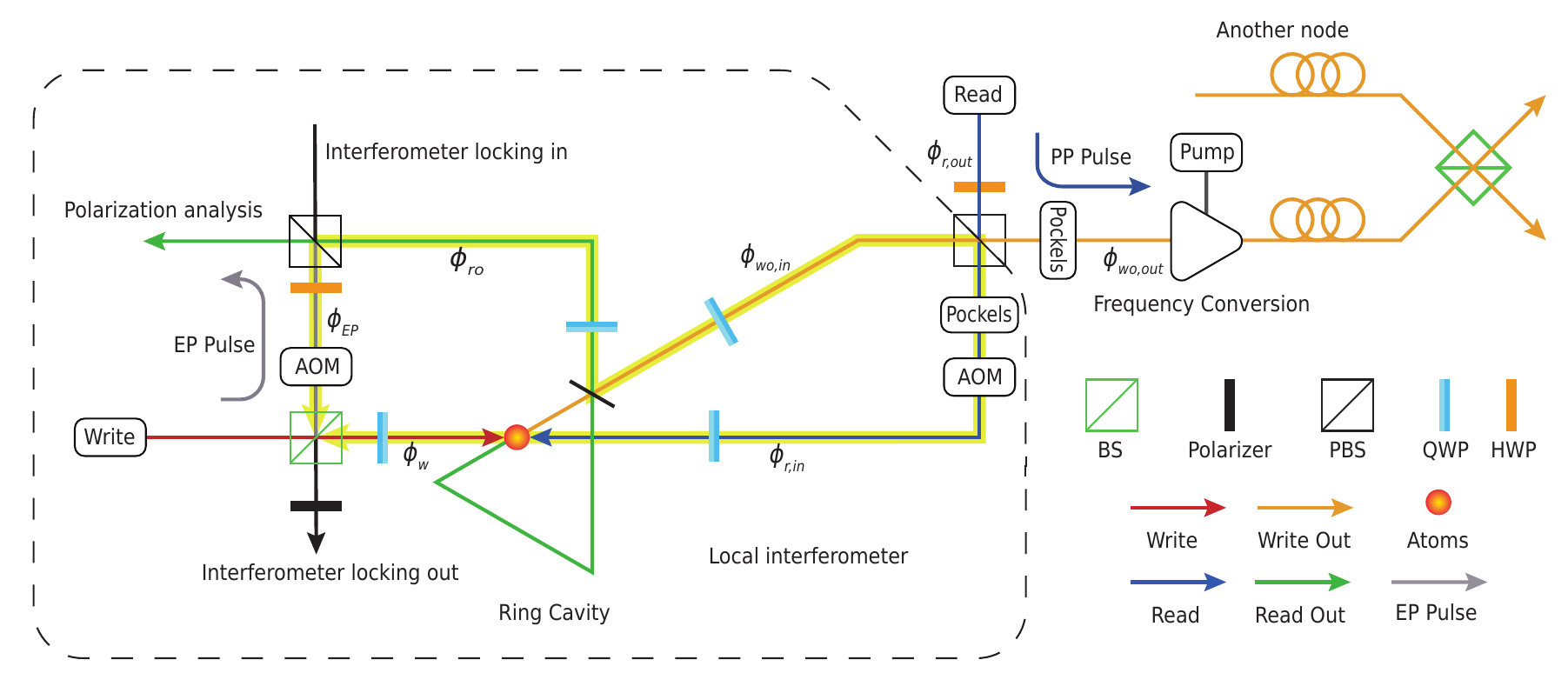}
	\caption{Detailed schematic of a quantum node. Two arms of a local interferometer are highlighted. PBS: polarizing beam splitter. BS: beam splitter. HWP: half-wave plate. QWP: quarter-wave plate. Pockels: Pockels cell. AOM: acoustic-optic modulator.}
	\label{fig:optical_paths_phase}
\end{figure}

Here we give a complete description of the phase involved in the remote entanglement generation and verification. Fig.~\ref{fig:optical_paths_phase} depicts a detailed schematic of a quantum node. The notations used in the following derivation and their corresponding description are listed in Tab.~\ref{tab:symbols_of_phase}. 

\begin{table*}[ht]
	\centering
	\caption{A summary of notations and their descriptions in the phase evolution derivation. For the phase notations, we use the superscript A, B, and C for referring to the node Alice, Bob, and Charlie, respectively.}
	\begin{tabular}[t]{C{2cm} C{10cm}}
	\toprule
	Notations & Description \\
	\midrule
	$t_w/t_r$& The time for the write-in or read-out operation, respectively.\\
	\\
	$\varphi_w(t)$& The initial phase of the Write laser at time $t$.\\
	$\varphi_p(t)$& The initial phase of the Pump laser at time $t$.\\
	$\varphi_r(t)$& The initial phase of the Read laser at time $t$.\\
	\\
	$\phi_w$& The optical path phase of the Write pulse.\\
	$\phi_r$& The optical path phase of the Read pulse.\\
	$\phi_{r, in}$& The part of $\phi_{r}$ accumulated in the local interferometer.\\
	$\phi_{r, out}$& The part of $\phi_{r}$ accumulated outside the local interferometer.\\
	$\phi_{wo}$& The optical path phase of the write-out photon.\\
	$\phi_{wo, in}$& The part of $\phi_{wo}$ accumulated in the local interferometer.\\
	$\phi_{wo, out}$& The part of $\phi_{wo}$ accumulated outside the local interferometer.\\
	$\phi_{ro}$& The optical path phase of the read-out photon.\\
	$\phi_{\textrm{EP}}$& The optical path phase of the EP pulse.\\
	\\
	$\phi_{a}$& The phase accumulated during atomic storage.\\
	$\phi_{wr}$& The total phase of the write-out-read-out photon pair.\\
	$\phi_{\textrm{EOM}}$& The feedback phase of the EOM at the server node.\\
	$\phi_{lo}$& The local phase related to a single node.\\
	$\phi_{rm}$& The remote phase related to two nodes.\\
	$\phi_{\textrm{PME}}$& The phase of the PME state.\\
	\bottomrule
	\end{tabular}
	\label{tab:symbols_of_phase}
\end{table*}

We start from the local atom-photon entanglement pair generation. The interference and the detection of the write-out photons herald the entanglement of two atomic ensembles. Then the atomic excitations are retried as photons for entanglement verification. The retrieval and write-out photon detection processes are communitive, so we consider a delayed-choice experiment where the atomic state is retrieved immediately after its creation. In this case, the local entanglement between the write-out and the read-out photon mode reads
\begin{equation}
	\ket{00}+\sqrt{\chi}e^{i\phi_{wr}}\ket{1}_{ro}\ket{1}_{wo},
 \end{equation}
 where
 \begin{equation}
 \phi_{wr}=\varphi_w(t_w)+\varphi_p(t_w)+\varphi_r(t_r)+\phi_a+\phi_w+\phi_{wo}+\phi_r+\phi_{ro}. 
 \end{equation}
 $\varphi_f=\phi_w+\phi_{wo}$ correspond to the phase of the fiber channel in the main text. With a click event of an SNSPD, the remote entangled state reads
 \begin{equation}\label{eq:ent_ro}
	\ket{\psi_{ro}^\pm}=\left(e^{i\phi_{wr}^\textrm{A}+i\phi_{\textrm{EOM}}}\ket{10}_{\textrm{AB}}\pm e^{i\phi_{wr}^\textrm{B}}\ket{01}_{\textrm{AB}}\right)/\sqrt{2}.
 \end{equation}
Apart from the phase of photon pairs $\phi_{wr}^\textrm{A/B}$, the phase $\phi_\textrm{EOM}$ is involved because the phase feedback is applied before the remote entanglement generation.

During the entanglement verification, we combine an EP pulse with the read-out photons at each node. We generate the EP pulse from the Write laser (because of its frequency likeliness with the read-out photon). Hence the EP pulse inherits the initial phase of the Write laser. In the limit of weak power, we can write the joint state of two EP pulses in two nodes as follows,
 \begin{equation}\label{eq:ent_ep}
	\ket{00}_\textrm{AB}+\alpha e^{i\varphi_w^\textrm{A}(t_r)+i\phi_\textrm{EP}^\textrm{A}}\ket{10}_\textrm{AB}+\alpha e^{i\varphi_w^\textrm{B}(t_r)+i\phi_\textrm{EP}^B}\ket{01}_\textrm{AB}.
 \end{equation}
 We encode the read-out photon modes in Eq.~\ref{eq:ent_ro} in the horizontal polarization ($\ket{H}$), the EP pulse mode in Eq.~\ref{eq:ent_ep} in the vertical polarization mode ($\ket{V}$) and combine them using PBSs (shown in the Fig.~3A of the main text). By selecting one photon at each node, we obtain a polarization maximally entangled (PME) state
 \begin{equation}
	\ket{\psi_\textrm{PME}^\pm}=
	\left(e^{i[\phi_{wr}^\textrm{A}+i\phi_\textrm{EOM}+\varphi_w^\textrm{B}(t_r)+\phi_\textrm{EP}^\textrm{B}]}\ket{HV}_\textrm{AB}\pm e^{i[\phi_{wr}^\textrm{B}+\varphi_w^\textrm{A}(t_r)+\phi_\textrm{EP}^\textrm{A}]}\ket{VH}_\textrm{AB} \right)/\sqrt{2}.
 \end{equation}
We define $\phi_\textrm{PME}$ as the phase of the PME state,
\begin{equation}
	\begin{split}
	\phi_\textrm{PME}=&[\phi_\textrm{EOM}+\phi_{wr}^\textrm{A}-\varphi_w^\textrm{A}(t_r)-\phi_\textrm{EP}^\textrm{A}]-[\phi_{wr}^\textrm{B}-\varphi^\textrm{B}_w(t_r) - \phi_\textrm{EP}^\textrm{B}]\\
	=&\phi_{lo}^\textrm{A}-\phi_{lo}^\textrm{B}+\phi_{rm},
	\end{split}
 \end{equation}
where we divide the total phase into the local phase at nodes A and B ($\phi^\textrm{A/B}_{lo}$) and the remote phase between two nodes ($\phi_{rm}$). They are given by:
\begin{equation}\label{eq:local_phase}
	\begin{split}
		\phi_{lo}^\textrm{A/B}=
		 &\phi_a^\textrm{A/B}+
		\left[\phi_w^\textrm{A/B}+\phi_{wo,in}^\textrm{A/B}+\phi_{r,in}^\textrm{A/B}+\phi_{ro}^\textrm{A/B}-\phi_{EP}^\textrm{A/B}\right]\\
		&+\left[\varphi_w^\textrm{A/B}(t_w)+\varphi_{r}^\textrm{A/B}(t_r)-\varphi_{w}^\textrm{A/B}(t_r)-\varphi_{r}^\textrm{A/B}(t_w)\right],
	\end{split}
\end{equation}
\begin{equation}\label{eq:remote_phase}
	\phi_{rm}=\phi_\textrm{EOM}+[\phi_{wo,out}^A+\phi_{r,out}^A+\varphi_r^A(t_w)+\varphi_p^\textrm{A}(t_w)]-[\phi_{wo,out}^B+\phi_{r,out}^B+\varphi_r^B(t_w)+\varphi_p^\textrm{B}(t_w)].
\end{equation}
In the expression of $\phi^\textrm{A/B}_{lo}$ (Eq.~\ref{eq:local_phase}), the first term $\phi_a^\textrm{A/B}$ stands for the phase during the storage in the atomic ensembles. The first bracket represents the phase difference of some optical paths, constituting an interferometer highlighted in Fig.~\ref{fig:optical_paths_phase}. The second bracket indicates the phase inherited from the lasers at different times. 

To keep $\phi_a^\textrm{A/B}$ stable in the 100~$\upmu$s time scale, we reduce the Zeeman energy level fluctuation of atoms by securing a constant magnetic environment (see Ref.~\cite{luo2022sm}) and transferring the spin wave to clock transition to reduce the sensitivity to the magnetic field. The stable laser phase is guaranteed by locking lasers to a stable and narrow linewidth reference cavity. To stabilize the local interferometer, we inject a locking beam during the MOT loading phase and apply the feedback onto a mirror glued with a piezo-electric transducer. The acousto-optic modulators (AOMs) on the optical paths function as frequency shifters. One shifts the PP pulse to the write-out photon's frequency during the phase probing phase, and the other shifts the EP pulse to the read-out photon's frequency during the entanglement verification phase.

We lock the remote phase $\phi_{rm}$ with the weak-field remote phase stabilization methods. For practical reasons, we add an addition $\pi/2$ in the feedback (i.e. $\phi_{rm}=\pi/2$), which implies,
\begin{equation}\label{eq:PME_phi}
	\begin{cases}
		\phi_\textrm{PME}^\textrm{A-B}&=\phi_{lo}^\textrm{A}-\phi_{lo}^\textrm{B}+\frac{\pi}{2}\\
		\phi_\textrm{PME}^\textrm{A-C}&=\phi_{lo}^\textrm{A}-\phi_{lo}^\textrm{C}+\frac{\pi}{2}\\
		\phi_\textrm{PME}^\textrm{C-B}&=\phi_{lo}^\textrm{C}-\phi_{lo}^\textrm{B}+\frac{\pi}{2}
	\end{cases}.
\end{equation}
By adding three equations in Eq.~\ref{eq:PME_phi}, we get:
\begin{equation}
	\phi_\textrm{PME}^\textrm{A-C}+\phi_\textrm{PME}^\textrm{C-B}-\phi_\textrm{PME}^\textrm{A-B}=\frac{\pi}{2},
\end{equation}
which explains the $\pi/2$ phase delay between $\ket{\Psi^\pm}_\textrm{BC}$ and $\ket{\Psi^\pm}_\textrm{AC}$ in Fig.~4A and 4B in the main text. 

\section{The model of entanglement verification}

Using the EP pulses, we can transfer the Fock state entanglement to PME states, which makes entanglement verification much simpler. Thus, we can use entanglement fidelity to quantify the quality of entanglement. Here, we give the form of entanglement fidelity and show the limitation under our experimental imperfections.

\subsection{The model of the state and the measurement}
The click of an SNSPD heralds the entanglement of two atomic ensembles. We consider the atomic state being retrieved as read-out photons. The density matrix of the joint read-out state under the Fock basis $\{\ket{00}, \ket{01}, \ket{10}, \ket{11}\}$ reads~\cite{chou2005sm}
\begin{equation}
	\rho_{ret}=\left(\begin{matrix}
		p_{00}& 0& 0&0\\
		0& p_{01}& d e^{i\phi}&0\\
		0& de^{-i\phi}& p_{10}&0\\
		0& 0& 0&p_{11}\\
	\end{matrix}
	\right),
	\label{eq: density matrix}
\end{equation}
where $p_{ij}$ represents $i$ photons at Alice and $i$ photons at Bob, and $d$ represents the coherence term. To certify the remote entanglement, we generate an EP pulse at each quantum node, which is essentially a weak coherent pulse resembling the read-out mode in all degree-of-freedoms. The density matrix of the joint EP state is
\begin{equation}
	\rho_\textrm{EP}=\ket{\alpha e^{i\theta_A}}\bra{\alpha e^{i\theta_A}}\otimes\ket{\alpha e^{i\theta_B}}\bra{\alpha e^{i\theta_B}},
	\label{eq: density matrix 2}
\end{equation}
where $\theta_{A}$ ($\theta_{B}$) represents the phase of the EP pulses at Alice (Bob). We encode the read-out photon onto the horizontal polarization, the EP pulse onto the vertical polarization and mix them using a PBS at each node. The state after mixing reads
\begin{equation}
	\rho_{\textrm{EP},V}\otimes \rho_{ret,H}.
	\label{eq:EPxRE}
\end{equation}
After mixing, we perform the polarization projection measurement. We register the events where only one detector clicks at each node. When measuring at the $\ket{H}/\ket{V}$ basis, the four associated observables reads
\begin{equation}
	\begin{cases}
		\hat{W}_{H,H}=\ket{0}\bra{0}_{A,V}\otimes(I-\ket{0}\bra{0})_{A,H}\otimes(I-\ket{0}\bra{0})_{B,H}\otimes\ket{0}\bra{0}_{B,V}\\
		\hat{W}_{V,H}=(I-\ket{0}\bra{0})_{A,V}\otimes\ket{0}\bra{0}_{A,H}\otimes(I-\ket{0}\bra{0})_{B,H}\otimes\ket{0}\bra{0}_{B,V}\\
		\hat{W}_{H,V}=\ket{0}\bra{0}_{A,V}\otimes(I-\ket{0}\bra{0})_{A,H}\otimes\ket{0}\bra{0}_{B,H}\otimes(I-\ket{0}\bra{0})_{B,V}\\
		\hat{W}_{V,V}=(I-\ket{0}\bra{0})_{A,V}\otimes\ket{0}\bra{0}_{A,H}\otimes\ket{0}\bra{0}_{B,H}\otimes(I-\ket{0}\bra{0})_{B,V}\\
	\end{cases},
\end{equation}
 where $\hat{W}_{m,n}$ represents the detection of a photon with $m$ polarization at node A and a photon with $n$ polarization at node B with ($m,n$ takes $H$ or $V$). 
 
\subsection{PME states and their fidelity estimation}
We define the state in Eq.~\ref{eq:EPxRE} when choosing one photon at each node as a polarization maximally entangled (PME) state. The correlator $ZZ$ of the PME state can be calculated as
\begin{equation}
	\begin{split}
	\braket{ZZ}=&\frac{Tr(\rho(\hat{W}_{V,V}+\hat{W}_{H,H}-\hat{W}_{V,H}-\hat{W}_{H,V}))}{Tr(\rho(\hat{W}_{V,V}+\hat{W}_{H,H}+\hat{W}_{V,H}+\hat{W}_{H,V}))}\\
	=&-\frac{(p_{01}+p_{10})(1-e^{-n})e^{-n}-p_{00}(1-e^{-n})^2-p_{11}e^{-2n}}{(p_{01}+p_{10})(1-e^{-n})e^{-n}+p_{00}(1-e^{-n})^2+p_{11}e^{-2n}}.
	\end{split}
\end{equation}
Similarly, the correlator $XX$ can be calculated as,
\begin{equation}
	\begin{split}
	\braket{XX}=&\frac{Tr(\rho(\hat{W}_{A,A}+\hat{W}_{D,D}-\hat{W}_{A,D}-\hat{W}_{D,A}))}{Tr(\rho(\hat{W}_{A,A}+\hat{W}_{D,D}+\hat{W}_{A,D}+\hat{W}_{D,A}))}\\
	=&\frac{2dn\cos(\phi+\theta_{A}-\theta_{B})}{4p_{00}(1-e^{-n/2})^2+2(p_{01}+p_{10})(1+\frac{n}{2})(1-e^{-n/2})+p_{11}(1+\frac{n}{2})^2}.
	\label{eq: XY visibility}
	\end{split}
\end{equation}
The observables $\hat{W}_{A/D,D/A}$ represent the detection of a photon with $\ket{A/D}$ polarization at node A  and a photon with $\ket{D/A}$ polarization at node B. We get them using the basis transformation
\begin{equation}
	\hat{W}_{A/D,D/A}=\hat{M}^\dagger(\hat{a}_{A,H},\hat{a}_{A,V})\hat{M}^\dagger(\hat{a}_{B,H},\hat{a}_{B,V})\hat{W}_{H/V,V/H}\hat{M}(\hat{a}_{A,H},\hat{a}_{A,V})\hat{M}(\hat{a}_{B,H},\hat{a}_{B,V}),
\end{equation}
where $\hat{M}(\hat{a}_{H},\hat{a}_{V})=e^{\pi/4(\hat{a}_H^\dagger\hat{a}_V-\hat{a}_V^\dagger\hat{a}_H)}$ is the mode mixing operator. The correlator $YY$ has the same form as $XX$, so we omit it here.

By fixing the phase of the EP pulses, we target the PME states to two Bell state $\ket{\Psi^\pm}$, with the sign $\pm$ depending on the phase of the initial atomic entanglement. We can estimate the fidelity of the PME states to their target Bell state as
\begin{equation}
	\mathcal{F}(\ket{\Psi^\pm})=\frac{\pm\langle XX\rangle\pm\langle YY\rangle-\langle ZZ\rangle+1}{4}.
\end{equation}
Substituting the expressions of the correlators we learn that the fidelity is a function of the EP state intensity $n$. In our experiments, we set $n\approx \sqrt{p_{11}/p_{00}}\approx 3\%\ll 1$ to maximize the fidelity. After ignoring the higher-order terms, $o(n)$, the fidelity is given by:
\begin{equation}
	\mathcal{F}(\ket{\Psi^\pm})\approx\frac{1+V}{2(1+\beta)},
	\label{eq: fidelity}
\end{equation}
where $V=2d/(p_{01}+p_{10})$ and $\beta=2\sqrt{p_{00}p_{11}}/(p_{01}+p_{10})$. The former describes the coherence between $p_{01}$ and $p_{10}$, which can also be understood as the interference visibility when we combine two read-out modes with a beamsplitter, and the latter describes the contribution of high order excitations.

To have more insight of the PME fidelity, we associate the fidelity expression with the atomic state. We define $d_{ij}$ as $i$ atomic excitations at node A and $j$ at node B upon an SNSPD heralding the remote entanglement between two atomic ensembles. We can associate $d_{ij}$ and $p_{ij}$ using the retrieval efficiency $\eta_r$,
\begin{equation}
	\begin{cases}
		p_{00}\approx d_{00}+(p_{10}+p_{01})(1-\eta_r)\\
		p_{10}\approx d_{10}\eta_r\\
		p_{01}\approx d_{01}\eta_r\\
		p_{11}\approx d_{11}\eta_r^2\\
	\end{cases},
\end{equation}
where we neglected the small contribution from $d_{11}$ on $p_{00},\ p_{01},\ p_{10}$. The non-zero part of $d_{00}$ comes from the incorrect heralding of the SNSPD, such as a dark count. $d_{11}$ is given by the higher-order atomic excitations. We estimate them as follows,
\begin{equation}
	\begin{cases}
		d_{00}\approx (d_{10}+d_{01})/\textrm{SNR}\\
		d_{11}\approx (d_{10}+d_{01})\chi\\
	\end{cases},
\end{equation}
where $\textrm{SNR}$ is the signal-to-noise ratio of write-out photon, and $\chi$ is the internal excitation probability. Then we can get:
\begin{equation}
	\beta\approx 2\sqrt{(1/\textrm{SNR}+1-\eta_r)\chi}.
	\label{eq: beta}
\end{equation}
Substituting it into Eq. \ref{eq: fidelity}, we obtain:
\begin{equation}
	\mathcal{F}(\ket{\Psi^\pm})\approx\frac{1+V}{2(1+2\sqrt{(1/\textrm{SNR}+1-\eta_r)\chi})}.
\end{equation}
From the above equation, retrieval efficiency does not significantly affect fidelity, while internal excitation probability and SNR play a significant role.

\subsection{Concurrence}
Concurrence $\mathcal{C}=max\{0,2(d-\sqrt{p_{00}p_{11}})\}>0$ is a widely used entanglement criterion for Fock state entanglement. Due to its monotonicity under local operation and classical communication (LOCC), $\mathcal{C}>0$ for the read-out photons verifies the entanglement of the atomic ensembles at two nodes. The key to estimating the entanglement is the coherence term $d$ measurement, which is usually done by interfering with the read-out photons and measuring the visibility \cite{chou2005sm, yu2020asm, lago-rivera2021sm}. Here, we show that we can give a lower bound on $d$ according to the visibility in the $X/Y$ basis, and thus we can give a lower bound on the concurrence.

From Eq. \ref{eq: XY visibility}, we can get:
\begin{equation}
	|\langle XX\rangle|=|\langle YY\rangle|<\frac{2dn}{4p_{00}\frac{n^2}{4(1+n/2)^2}+(p_{01}+p_{10})n+p_{11}(1+\frac{n}{2})^2}\le \frac{V}{1+\beta}=\frac{2d}{p_{01}+p_{10}+2\sqrt{p_{11}p_{00}}}.
	\label{eq: XY bound}
\end{equation}
Then, we can get:
\begin{equation}
	2d>E_{X/Y}(p_{01}+p_{10}+2\sqrt{p_{00}p_{11}}),
\end{equation}
where $E_{X/Y}$ is the absolute value of the correlator XX/YY. Then, the concurrence is  given by:
\begin{equation}
	\mathcal{C}>\mathcal{C}_l=max\{0,E_{X/Y}(p_{01}+p_{10}+2\sqrt{p_{00}p_{11}})-2\sqrt{p_{00}p_{11}}\}.
\end{equation}
According to the data measured in $Z$ basis, we can get the photon statistics $p_{ij}$ of the read-out photons. Furthermore, with $E_{X/Y}$ (we use the mean value of $E_X$ and $E_Y$ to calculate $d$) we can give a lower bound on the concurrence $\mathcal{C}$. 

\subsection{Photon transmission losses and non-unit detection efficiency}
The above calculations are based on unit detection efficiency, and the losses during the photon transmission are not considered. 
This section will show that these two factors do not affect the above conclusion.The non-unit transmission efficiency $\eta$ of a mode $\hat{a}$ can be modeled by a BS with transmission of $\eta=\cos(\kappa)^2$, whose operator is:
\begin{equation}
	\hat{L}(\hat{a})=e^{\kappa(\hat{a}^\dagger\hat{l}_a-\hat{l}_a^\dagger\hat{a})},
\end{equation}
where $\hat{a}$ is the mode we care about, and $\hat{l}_a$ is an auxiliary mode that models the losses of photons in $\hat{a}$ mode. This loss operator, on the one hand, describes the losses due to all kinds of optical devices; on the other hand, it describes the non-unit detection efficiency. The partial density matrix of the total density matrix \ref{eq: density matrix 2} at one node is a mixture of general two-mode quantum state, which is given by:
\begin{equation}
	\ket{\psi}=\sum_{i,j}c_{i,j}\ket{i}\ket{j},
\end{equation}
where $\ket{i}$ is the Fock state basis. The general two-mode quantum state can be also transferred to the coherent state basis with the completeness relation:
\begin{equation}
	I=\frac{1}{\pi}\int\ket{\alpha}\bra{\alpha}d^2\alpha.
\end{equation}
Then, it can be written in the form:
\begin{equation}
	\ket{\psi}=\iint c(\alpha,\beta)\ket{\alpha}\ket{\beta}d^2\alpha d^2\beta.
\end{equation}
In coherent state basis, it's easy to check the following equation:
\begin{equation}
	\hat{M}^\dagger(\hat{l}_a,\hat{l}_b)\hat{L}(\hat{a})\hat{L}(\hat{b})\hat{M}(\hat{a},\hat{b})\ket{\psi}=\hat{M}(\hat{a},\hat{b})\hat{L}(\hat{a})\hat{L}(\hat{b})\ket{\psi},
\end{equation}
which means that after tracing all the auxiliary modes $\hat{l}_a,\ \hat{l}_b$, we can exchange the position of the loss operator $\hat{L}(\hat{a})$ and the mode mixing operator $\hat{M}(\hat{a},\hat{b})$. So, in the $X/Y$ basis the non-unit detection efficiency is equivalent to the losses before the mode mixing operation. This equivalence in the $Z$ basis is trivial.

Now, we can concentrate on the losses of EP pulses and the read-out photons. For the former one, we can get:
\begin{equation}
	Tr_{\hat{l}_a}[\hat{L}(\hat{a})\ket{\alpha}]=\ket{\sqrt{\eta}\alpha},
\end{equation}
which shows that losses only change the amplitude of the EP pulses. For the latter, the losses can be regarded as a decrease in the retrieval efficiency, which does not change the form of the density matrix \ref{eq: density matrix}. Thus, by adjusting the amplitude of the EP pulses and redefining the density matrix of the read-out photons(including all the losses and the limited detection efficiency), the conclusion derived in the ideal cases still works in the realistic experimental parameters.

\section{Data analysis}
 \subsection{The imperfection of $\langle ZZ \rangle$}
 We have given the form of the visibility in $Z$ basis in Eq. \ref{eq: XY visibility}. We find that it is subject to the  following inequality:
\begin{equation}
	-\langle ZZ \rangle\le \frac{1-\beta}{1+\beta},
\end{equation}
where the equality holds under the condition $e^{n}-1=\sqrt{p_{11}/p_{00}}$, which is approximately satisfied in our experiments. Then we can compare the experimental results $V_E$ with  $(1-\beta)/(1+\beta)$, where $\beta$ can be given by definition $V_\beta$ or estimated from the experimental parameters $V_P$ according to Eq. \ref{eq: beta}. The results are given in Tab. \ref{tab:HV imperfection}:
\begin{table*}[!htbp]
	\caption{Experimental and  theoretical  compression of visibility in  Z basis.} 
	\scriptsize
	\label{tab:HV imperfection}
	\renewcommand\arraystretch{1.2}
	\begin{ruledtabular}
	\begin{tabular}{C{2.5cm} R{1cm} C{2.1cm} C{2.1cm} C{2.1cm} }
		& &  $\bm{V_E}$  &  $\bm{V_\beta}$  &  $\bm{V_P}$ \\
		\hline
		\multirow{2}*{\textbf{Storage time (ST)} $\bm{5 \mu s}$}& $\bm{\psi^-}$ & $0.60\pm 0.07$ & $0.631\pm0.021$ & $0.629\pm0.016$\\
		&$\bm{\psi^+}$ & $0.67\pm 0.07$ & $0.661\pm0.022$& $0.619\pm0.016$ \\
		\specialrule{0em}{3pt}{3pt}
		\multirow{2}*{\textbf{Storage time (ST)} $\bm{107 \mu s}$}&$\bm{\psi^-}$  & $0.56\pm 0.09$ & $0.551\pm0.032$  & $0.561\pm0.024$ \\
		&$\bm{\psi^+}$  & $0.69\pm 0.09$ &$0.605\pm0.035$ & $0.548\pm0.025$ \\
		\specialrule{0em}{3pt}{3pt}
		\multirow{2}*{\textbf{A-B}}& $\bm{\psi^-}$ & $0.71\pm 0.09$ & $0.674\pm0.029$ & $0.671\pm0.021$ \\
		&$\bm{\psi^+}$ & $0.72\pm 0.09$ & $0.685\pm0.029$ & $0.665\pm0.021$ \\
		\specialrule{0em}{3pt}{3pt}
		\multirow{2}*{\textbf{A-C}}&$\bm{\psi^-}$  & $0.48\pm 0.08$ & $0.513\pm0.027$ & $0.510\pm0.021$ \\
		&$\bm{\psi^+}$  & $0.53\pm 0.008$ &  $0.577\pm0.028$  & $0.511\pm0.021$ \\
		\specialrule{0em}{3pt}{3pt}
		\multirow{2}*{\textbf{B-C}}&$\bm{\psi^-}$ & $0.55\pm 0.09$ & $0.586\pm0.028$ & $0.541\pm0.021$ \\
		&$\bm{\psi^+}$ & $0.49\pm 0.09$ & $0.555\pm0.028$ & $0.535\pm0.021$ \\
	\end{tabular}
\end{ruledtabular}
\end{table*}

We can see that the experimental results $V_E$ agree well with the estimated value $V_P$.

 \subsection{The imperfection of $V$}
In Eq.~\ref{eq: XY bound}, we have shown that the visibility $\langle XX/YY \rangle$ is bounded by $V/(1+\beta)$.
The parameter $\beta$ captures the imperfection due to the high-order excitations, and the left part $V$, defined as $2d/(p_{01}+p_{10})$, describes the coherence of the Fock state entanglement. It can also be interpreted as the interference visibility of the read-out photons when two read-modes are combined with a BS. Here, we focus on several experimental imperfections that reduce the visibility $V$, and we use $V_M$ to represent the maximum visibility under a particular factor.

\subsubsection{High order excitations}

The photon counts distribution $p_{01},\ p_{10}$ contain the contribution of high order excitations, which reduce the interference visibility.  Considering the balanced retrieval efficiency $\eta_r$  and the atomic excitation distribution $d_{10},\ d_{01},\ d_{11},\ d_{20},\ d_{02}$,  the photon count distribution $p_{10},\ p_{01}$ can be given by:
\begin{equation}
	\begin{cases}
		p_{10}=d_{10}\eta_r+d_{11}\eta_r(1-\eta_r)+d_{20}(1-(1-\eta_r)^2)\\
		p_{01}=d_{01}\eta_r+d_{11}\eta_r(1-\eta_r)+d_{02}(1-(1-\eta_r)^2)\\
	\end{cases},
\end{equation}
where the high order excitation terms are estimated as $d_{11}=d_{20}=d_{02}=\chi(d_{10}+d_{01})$. The maximum visibility is given by:
\begin{equation}
	V_M=\frac{d_{10}\eta_r+d_{01}\eta_r}{p_{01}+p_{10}}=\frac{1}{1+2\chi(3-2\eta_r)}.
\end{equation}
With the mean retrieval efficiency $\eta_r$ and the internal excitation probability $\chi$ substituted into the above equation, we get the maximum visibility $V_M$, as shown in Tab. \ref{tab:high order excitations on V}:

\begin{table*}[!htbp]
	\caption{$V_m$ under high order excitations}
	\label{tab:high order excitations on V}
	\renewcommand\arraystretch{1.2}
	\begin{ruledtabular}
	\begin{tabular}{cccccc}
		& $\bm{ST\  5\mu s}$ &  $\bm{ST \ 107\mu s}$ & $\bm{A-B}$& $\bm{A-C}$ & $\bm{B-C}$\\
		\hline
		$\bm{V_M}$ &$0.935\pm0.009$&$0.910\pm 0.016$&$0.954\pm0.009$&$0.894\pm0.014$&$0.909\pm0.013$\\
	\end{tabular}
\end{ruledtabular}
\end{table*}

\subsubsection{The indistinguishability of the read-out photons and the EP pulses}

We use Hong-Ou-Mandel (HOM) interference to estimate the indistinguishability of the read-out photons and the EP pulses. The unnormalized density matrix of the read-out photons and the phase-randomized EP pulses can be given by:
\begin{equation}
	\begin{split}
		\rho_{ro}&=\ket{0}\bra{0}+p_{ro}\ket{1}\bra{1}+p_{ro}^{(2)}\ket{2}\bra{2},\\
		\rho_{EP}&=\ket{0}\bra{0}+p_{EP}\ket{1}\bra{1}+p_{EP}^{(2)}\ket{2}\bra{2},
	\end{split}
\end{equation}
where $p_{ro/EP}$ are the average photon numbers of read-out photons and the EP pulses, $p_{ro/EP}^{(2)}$ represent the probability of two-photon term, and higher order terms are ignored. In ideal cases, $g^{(2)}$ of the read-out photons is given by 2, and $g^{(2)}$  of the EP pulses is given by 1.  After mixing two fields with a BS, we can get the coincidence probability:
\begin{equation}
	p_{12}=\frac{p_{ro}^2}{2}+\frac{p_{EP}^2}{4}+\frac{p_{ro}p_{EP}}{2}(1-\eta).
\end{equation}
The first and second terms come from the two-photon component in the read-out photons and EP pulses, and the last term comes from the single-photon component of both fields. If they are completely indistinguishable, the last term disappears, and $\eta$ represents the indistinguishability of two fields \cite{li2013sm}. When both fields are on, $g^{(2)}$ is given by:
\begin{equation}
	g^{(2)}=\frac{\frac{p_{ro}^2}{2}+\frac{p_{EP}^2}{4}+\frac{p_{ro}p_{EP}}{2}(1-\eta)}{(\frac{p_{ro}}{2}+\frac{p_{EP}}{2})^2}.
	\label{eq: g2}
\end{equation}
When $\eta=1,p_{EP}=2p_{ro}$, $g^{(2)}$ reaches its minimum value of  $2/3$. We get $\eta$ by setting $p_{ro}\approx 2\%,p_{EP}\approx 4\%$, and then measuring $g^{(2)}$. The measurement results of three nodes are listed in Tab. \ref{tab:Indistinguishability of read-out photons and coherent pulses}.
\begin{table*}[!htbp]
	\caption{$\bm{g^{(2)}}$ and $\bm{\eta}$ of three nodes}
	\label{tab:Indistinguishability of read-out photons and coherent pulses}
	\renewcommand\arraystretch{1.2}
	\begin{ruledtabular}
	\begin{tabular}{cccc}
		& $\bm{A}$ &  $\bm{B}$ & $\bm{C}$ \\
		\hline
		$\bm{\ \ g^{(2)}}$ &$0.716\pm0.019$&$0.717\pm 0.014$&$0.700\pm0.028$\\
		$\bm{\eta}$ &$0.90\pm0.04$&$0.90\pm0.04$&$0.95\pm0.08$\\
	\end{tabular}
\end{ruledtabular}
\end{table*}

 The indistinguishability of the read-out photons and the EP pulses will affect the visibility $V$.  We use the following polarization entangled state to model this imperfection:
\begin{equation}
	\frac{1}{\sqrt{2}}\ket{H}\ket{\phi_1}\ket{V}(\sqrt{\eta_2}\ket{\phi_2}+\sqrt{1-\eta_2}\ket{\phi_2^\perp})+\frac{1}{\sqrt{2}}\ket{V}(\sqrt{\eta_1}\ket{\phi_1}+\sqrt{1-\eta_1}\ket{\phi_1^\perp})\ket{H}\ket{\phi_2},
\end{equation}
 where $\ket{\phi}$ is the other degree of freedom of photons,$\braket{\phi_{1/2}|\phi_{1/2}^\perp}=0$, $\eta_{1/2}$ represent the indistinguishability parameter at two nodes. In the superposition basis, the visibility is given by  $\sqrt{\eta_1\eta_2}$. This value describes the decrease of $V$ due to the mode mismatch, and related experimental results are given in Tab. \ref{tab: modes mismatch on V}
\begin{table*}[!htbp]
	\caption{$V_M$ under mode mismatch}
	\label{tab: modes mismatch on V}
	\renewcommand\arraystretch{1.2}
	\begin{ruledtabular}
	\begin{tabular}{cccccc}
		& $\bm{ST\  5\mu s}$ &  $\bm{ST \ 107\mu s}$ & $\bm{A-B}$& $\bm{A-C}$ & $\bm{B-C}$\\
		\hline
		$\bm{V_M}$ &$0.900\pm0.028$&$0.900\pm 0.028$&$0.900\pm0.028$&$0.92\pm0.04$&$0.92\pm0.04$\\
	\end{tabular}
\end{ruledtabular}
\end{table*}

\subsubsection{The imbalance between $p_{01}$ and $p_{10}$ }
The imbalance  of $p_{01}$ and $p_{10}$, which stems from the imbalance of the internal excitation probability $\chi$, the transmission efficiency of write-out photons $\eta_w$ and the retrieval efficiency $\eta_r$,  will also lead to the decrease of $V$. We have given the density matrix of the read-out photons in Eq. \ref{eq: density matrix}. The positive semidefiniteness of the density matrix puts an upper bound on $d$:
\begin{equation}
	d\le \sqrt{p_{01}p_{10}}.
\end{equation}
With the definition of $V$, we get:
\begin{equation}
	V\le \frac{\sqrt{p_{01}p_{10}}}{\frac{p_{01}+p_{10}}{2}}.
\end{equation}
The above equation shows the limitation of $V$ due to the imbalance of $p_{10},\ p_{01}$, and the experimental results are given in Tab. \ref{tab: imbalance on V}:
\begin{table*}[!htbp]
	\caption{$V_M$ under the imbalance of $p_{10},\ p_{01}$}
	\label{tab: imbalance on V}
	\renewcommand\arraystretch{1.2}
	\begin{ruledtabular}
	\begin{tabular}{cccccc}
		& $\bm{ST\  5\mu s}$ &  $\bm{ST \ 107\mu s}$ & $\bm{A-B}$& $\bm{A-C}$ & $\bm{B-C}$\\
		\hline
		$\bm{V_M}$ &$1.000\pm0.0012$&$0.995\pm 0.015$&$1.000\pm0.014$&$0.997\pm0.014$&$0.994\pm0.0014$\\
	\end{tabular}
\end{ruledtabular}
\end{table*}

We can see that the imbalance of $p_{01},\ p_{10}$ decreases the visibility $V$ little.

\subsubsection{The phase instability}

The phase instability is the critical factor that reduces the visibility $V$. The polarization entanglement states with a phase uncertainty $\delta \theta$ can be written as: $\ket{\psi^\pm}=1/\sqrt{2}(\ket{HV}\pm e^{i\delta \theta}\ket{VH})$. Considering a distribution of $f(\delta \theta)$, the average visibility is given by:
\begin{equation}
	V_M=\int_{-\infty}^{+\infty}f(\delta \theta)\cos(\delta \theta) d\delta \theta.
\end{equation}
We assume that $\delta \theta$ follows a Gaussian probability distribution with a mean of 0 and  a standard deviation ($SD$) of $\sigma$.  Then, we will get:
\begin{equation}
	V_M=e^{-\sigma^2/2},
	\label{eq: SD and VM}
\end{equation}
 which shows that the $SD$ of the phase is the critical parameter we need to consider.

\begin{figure}[htbp]
	\centering
	\label{fig: remote phase locking}
	\includegraphics[scale=0.33]{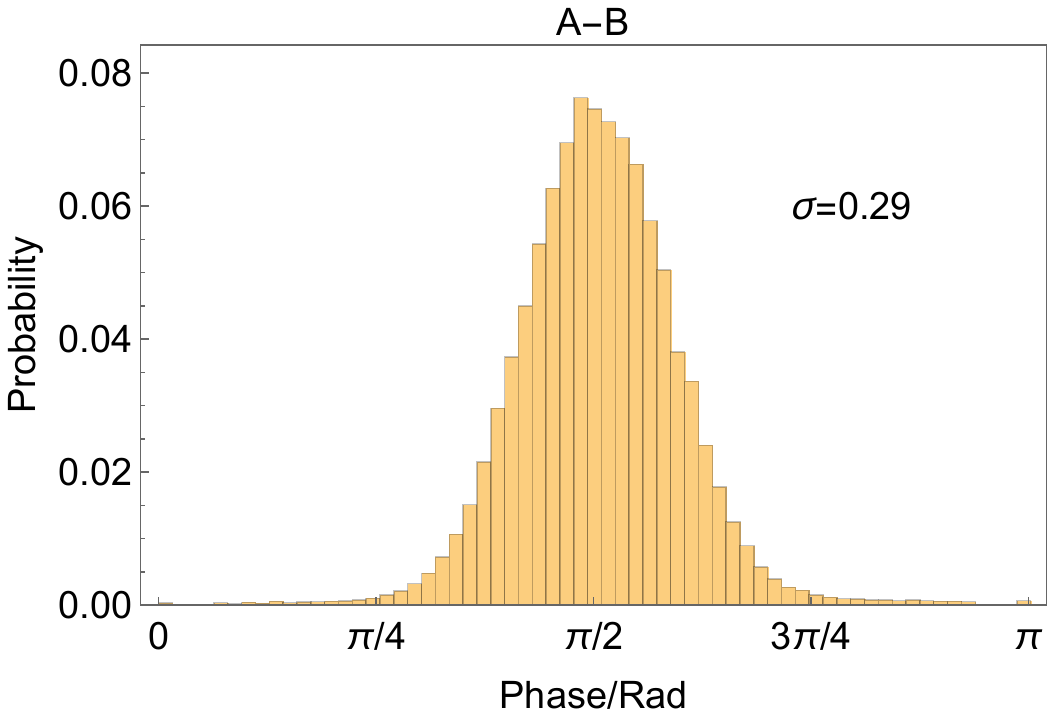}
	\includegraphics[scale=0.33]{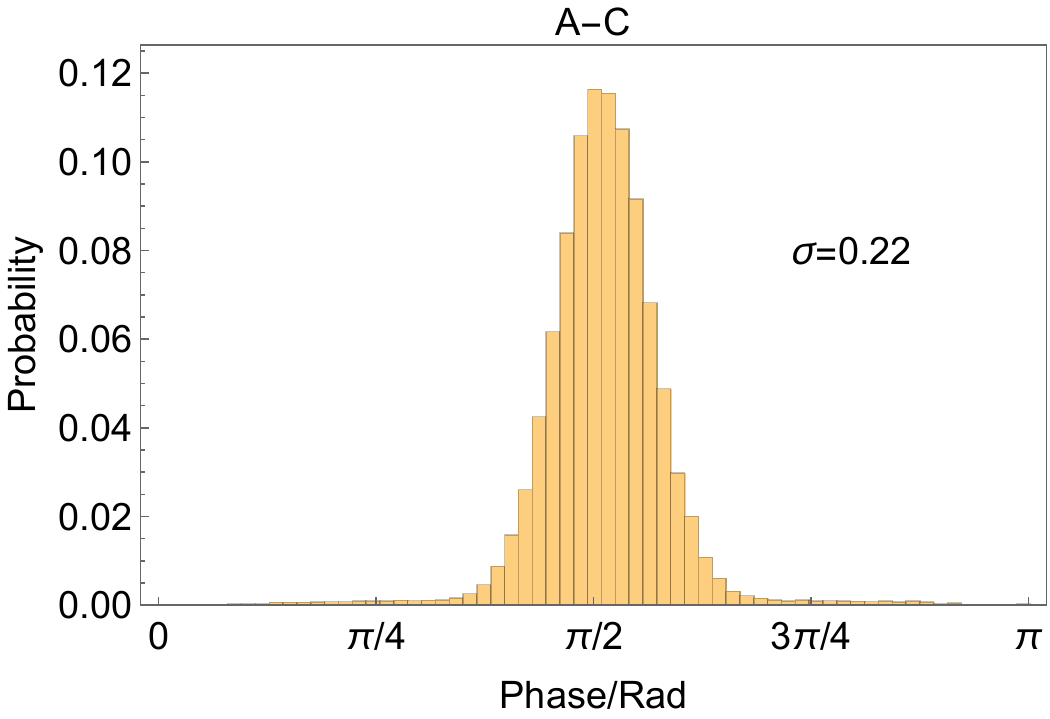}
	\includegraphics[scale=0.33]{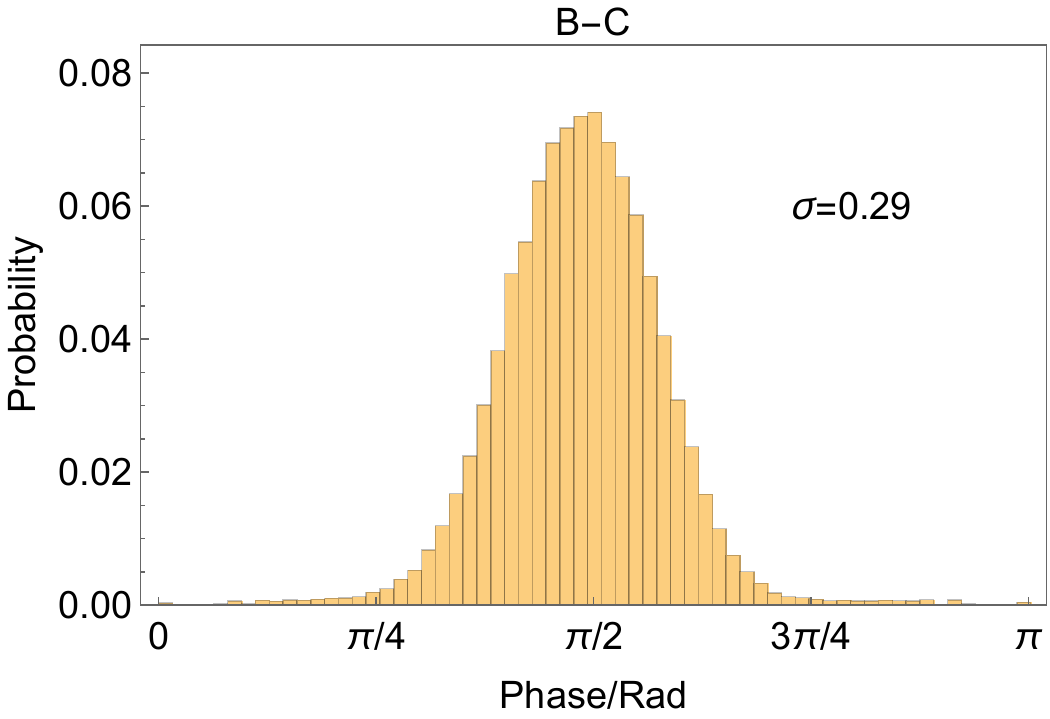}
	\caption{Remote phase locking precision estimation}
  \end{figure}

In the part $\uppercase\expandafter{\romannumeral2}$, we have given the phase of the entanglement states in $A-B$ connection: $\phi_{PME}^{A-B}=\phi_L^A-\phi_L^B+\phi_R^{A-B}$. Then the $SD$ is given by two local phases $SD[\phi_L^{A/B}]$ and one remote phase $SD[\phi_R^{A-B}]$. We estimate the $SD$ of the remote phase with the photon distribution of the phase probe laser after phase locking. The results  are given in Fig. \ref{fig: remote phase locking}. Due to the higher phase noise of the lasers in node B, the phase locking performance in the A-B/B-C connection is worse than in the A-C connection.
  
As for the local phase $SD[\phi_L]$, it consists of three parts: the memory phase $SD[\phi_a]$, the interferometer phase $SD[\varphi_{in}]$ and the laser phase $SD[\varphi_l]$. The memory phase is determined by the magnetic field at the atoms and the storage time:
\begin{equation}
	\phi_a=2\mu_Bg_FB\Delta t.
\end{equation}
The fluctuation of the magnetic field, which has a characteristic  frequency of 50 Hz, is mainly caused by the surrounding power supplies. The SD of the magnetic field in three nodes is given by $1.9mG,\ 0.8mG,\ 0.4mG$.
With the storage time of about $5\mu s$, we can get the SD of the memory phase, as shown in Tab. \ref{tab: phase instability}. As for the extended storage time experiments, the spin wave is transferred to clock transition, leading to three orders of magnitude reduction of the magnetic field sensitivity so that we can neglect the extra memory phase fluctuation due to longer storage time.

The phase instability coming from the local interferometers is estimated according to the interference signal after the phase locking, and the locking performance is also shown in Tab. \ref{tab: phase instability}. The significant difference in interferometer locking precision is due to  vibration noise in the environment.

\begin{figure}[htbp]
	\centering
	\includegraphics[width=.9\linewidth]{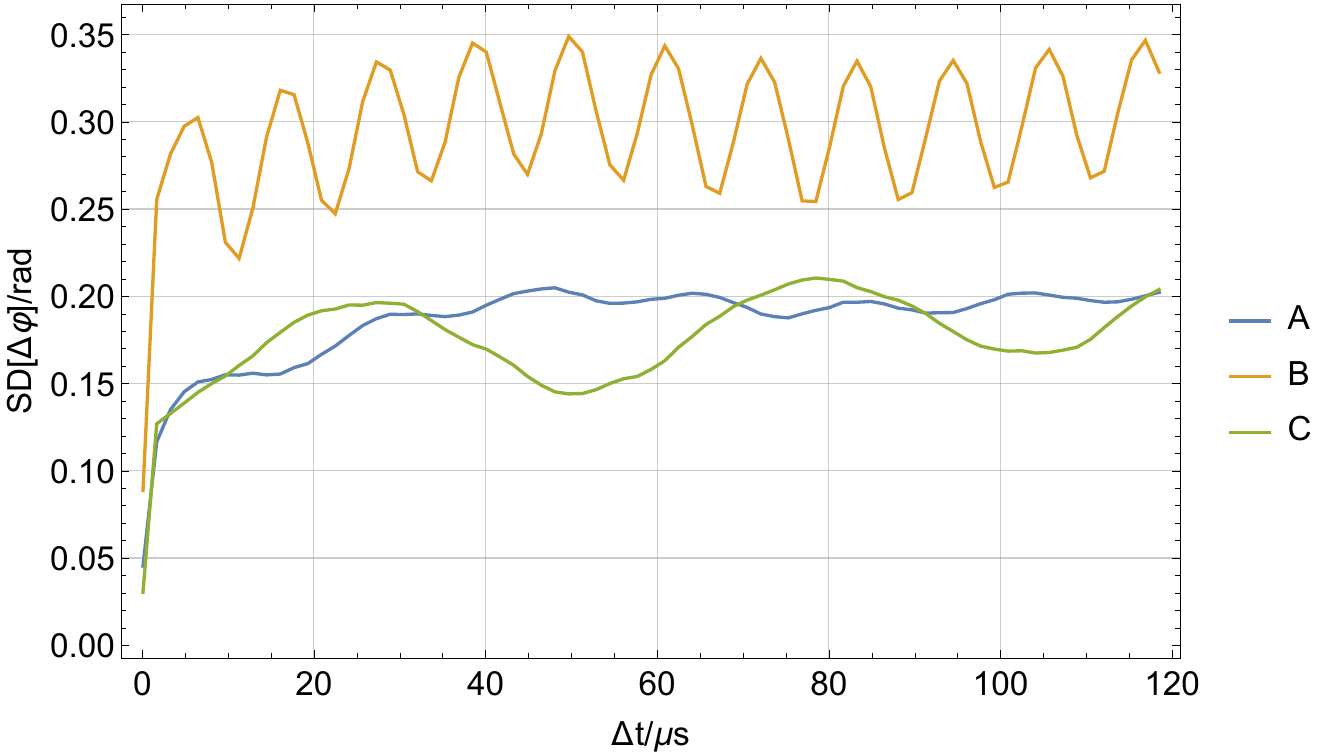}
	\caption{The standard deviation of the phase change in $\Delta t$}
	\label{fig:laser phase}
\end{figure}

The last part of the phase instability comes from the lasers. We estimate $SD[\varphi_l]$ according to the beat frequency of the write and read lasers, which are locked to an ultra-stable cavity and then phase-locked to each other. The beat signal at 6.8GHz is shifted to the DC range with a microwave signal source and a frequency mixer, and then recorded with an oscilloscope, whose bandwidth is 500 MHz.  In this way, we can measure the phase difference of write and read laser: $\varphi(t)=\varphi_w(t)-\varphi_r(t)$, and then we can estimate the phase instability of  lasers by $SD[\Delta\varphi(\Delta t)]$, where $\Delta\varphi(\Delta t)=\varphi(t+\Delta t)-\varphi(t)$, $\Delta t$ is the storage time. The measurement results are given in Fig. \ref{fig:laser phase} and Tab. \ref{tab: phase instability}. The phase oscillation shows the characteristic frequency noise of the lasers at each node. 
\begin{table*}[!htbp]
	\caption{The phase instability of memories, Interferometers and lasers.}
	\label{tab: phase instability}
	\renewcommand\arraystretch{1.2}
	\begin{ruledtabular}
	\begin{tabular}{cccccc}
		\textbf{Standard Deviation/rad}& \textbf{node} $\bm{A}$ & \textbf{node} $\bm{B}$ & \textbf{node} $\bm{C}$ \\
		\hline
		$\bm{SD[\phi_a]}$ &0.084&0.035&0.018\\
		$\bm{SD[\varphi_{in}]}$ &0.029&0.135&0.063\\
		$\bm{SD[\varphi_l]\  at \ 5 \mu s}$ &0.14&0.31&0.14\\
		$\bm{SD[\varphi_l] \ at\  107\mu s}$ &0.20&0.33&0.17\\
	\end{tabular}
\end{ruledtabular}
\end{table*}

With all the phase instability given above, the total $SD$ of the local phase  is given by:
\begin{equation}
	SD[\phi_L]=\sqrt{SD[\phi_a]^2+SD[\varphi_{in}]^2+SD[\varphi_l]^2}.
\end{equation}
Then the phase instability of the entanglement state of two nodes A and B is estimated as:
\begin{equation}
	SD[\phi_{PME}^{A-B}]=\sqrt{SD[\phi_L^A]^2+SD[\phi_L^B]^2+SD[\varphi_R^{A-B}]^2}.
\end{equation}
From Eq. \ref{eq: SD and VM} , we can evaluate the influence of the total phase instability on the visibility, and the results are given in Tab. \ref{tab: phase instability on V}

\begin{table*}[!htbp]
	\caption{$V_M$ under the phase instability}
	\label{tab: phase instability on V}
	\renewcommand\arraystretch{1.2}
	\begin{ruledtabular}
	\begin{tabular}{cccccc}
		& $\bm{ST\  5\mu s}$ &  $\bm{ST \ 107\mu s}$ & $\bm{A-B}$& $\bm{A-C}$ & $\bm{B-C}$\\
		\hline
		$\bm{V_M}$ &0.89&0.88&0.89&0.95&0.89\\
	\end{tabular}
\end{ruledtabular}
\end{table*}

\subsubsection{Summary}
\begin{table*}[!htbp]
	\caption{The comparison between the theoretical estimation of $V$ and the experimental results }
	\label{tab: theory and experiment of V}
	\renewcommand\arraystretch{1.2}
	\begin{ruledtabular}
	\begin{tabular}{cccccc}
		& $\bm{ST\  5\mu s}$ &  $\bm{ST \ 107\mu s}$ & $\bm{A-B}$& $\bm{A-C}$ & $\bm{B-C}$\\
		\hline
		\textbf{High order excitations} &$0.935\pm0.009$&$0.910\pm 0.016$&$0.954\pm0.009$&$0.894\pm0.014$&$0.909\pm0.013$\\
		\textbf{Mode mismatch} &$0.900\pm0.028$&$0.900\pm 0.028$&$0.900\pm0.028$&$0.92\pm0.04$&$0.92\pm0.04$\\
	    \textbf{Phase instability} &0.89&0.88&0.89&0.95&0.89\\
		\textbf{Imbalance of Read-out photons} &$1.000\pm0.0012$&$0.995\pm 0.015$&$1.000\pm0.014$&$0.997\pm0.014$&$0.994\pm0.0014$\\
		$\bm{V_{theory}}$ &$0.753\pm0.026$&$0.719\pm0.028$&$0.768\pm0.027$&$0.79\pm0.04$&$0.75\pm0.04$\\
		$\bm{V_{exp}}$ &$0.64\pm0.05$&$0.59\pm0.6$&$0.67\pm0.05$&$0.73\pm0.06$&$0.51\pm0.05$\\
	\end{tabular}
\end{ruledtabular}
\end{table*}

In the above sections, we have discussed several experimental imperfections that reduce the visibility $V$:

1. The high order excitations.

2. The indistinguishability of the read-out photons and the EP pulses.

3. The imbalance between $p_{01}$ and $p_{10}$.

4. The phase instability.

The main reducing factors are the high-order excitations, the mode mismatch, and the phase instability. At the same time, the imbalance of the read-out photons has little influence. For different quantum nodes, the laser phase noises at node B are much larger than the other two. Due to  more considerable transmission losses of the field deployed optical fibers between node C and the server node, we set a little higher internal excitation probability $\chi$  at node C to balance the transmission losses. These two factors result in worse visibility in the B-C connection. Here, all the maximum visibility $V_M$ under these experimental imperfections are summarized in Tab. \ref{tab: theory and experiment of V}, and we multiply them together as $V_{theory}$ to estimate their total contribution. The experimental results $V_{exp}$, calculated  according to its definition, are also given in Tab. \ref{tab: theory and experiment of V}.
The experimental results are lower than the theoretical estimate, which hints other uncaptured experimental imperfections. For example, the slow laser intensity drift of the local interferometer locking beam will lead to the drift of the phase locking point, thus reducing the visibility $V$.



\end{document}